\input harvmac
\input epsf
%
%
\noblackbox
\newcount\figno
\figno=0
\def\fig#1#2#3{
\par\begingroup\parindent=0pt\leftskip=1cm\rightskip=1cm\parindent=0pt
\baselineskip=11pt
\global\advance\figno by 1
\midinsert
\epsfxsize=#3
\centerline{\epsfbox{#2}}
\vskip -21pt
{\bf Fig.\ \the\figno: } #1\par
\endinsert\endgroup\par
}
\def\figlabel#1{\xdef#1{\the\figno}}
\def\encadremath#1{\vbox{\hrule\hbox{\vrule\kern8pt\vbox{\kern8pt
\hbox{$\displaystyle #1$}\kern8pt}
\kern8pt\vrule}\hrule}}
\def\ev#1{\langle#1\rangle}

\def\frac#1#2{{#1 \over #2}}

\def\semi{\subset\kern-1em\times\;}
\def\bar#1{\overline{#1}}
\def\sqr#1#2{{\vcenter{\vbox{\hrule height.#2pt
\hbox{\vrule width.#2pt height#1pt \kern#1pt \vrule width.#2pt}
\hrule height.#2pt}}}}

\def\CF{{\cal F}}

\def\CM{{\cal M}}                   \def\CN{{\cal N}}
\def\CO{{\cal O}}

\def\Tt{{\tilde{t}}}

\def\Qt{{\tilde{Q}}}
\def\Nt{{\tilde{N}}}

\def\Nh{{\hat{N}}}

\def\zb{\bar{z}}

\def\Tr{{\rm Tr}}

\def\Bracket#1{{\left\langle{#1}\right\rangle}}
\def\Pone{{\bf P}^1}
\def\Ztwo{{\bf Z}_2}

\def\mm#1{{\bf #1}}
\def\mg{\mm{g}}

\def\mF{\mm{F}}
\def\mA{\mm{A}}

\def\mN{\mm{N}}

\def\mR{\mm{R}}
\def\mS{\mm{S}}
\def\mZ{\mm{Z}}
\def\mPhi{{\mm{\Phi}}}

\def\Rt{{\tilde{R}}}
\def\St{{\tilde{S}}}
\def\Tt{{\tilde{T}}}

%
\lref\MatoneBX{
M.~Matone,
``The affine connection of supersymmetric SO(N)/Sp(N) theories,''
JHEP {\bf 0310}, 068 (2003)
[arXiv:hep-th/0307285].
}
\lref\FerrariJP{
F.~Ferrari,
``On exact superpotentials in confining vacua,''
Nucl.\ Phys.\ B {\bf 648}, 161 (2003)
[arXiv:hep-th/0210135].
}
\lref\AbbaspurHF{
R.~Abbaspur, A.~Imaanpur and S.~Parvizi,
``N = 2 SO(N) SYM theory from matrix model,''
JHEP {\bf 0307}, 043 (2003)
[arXiv:hep-th/0302083].
}
\lref\ItaKX{
H.~Ita, H.~Nieder and Y.~Oz,
``Perturbative computation of glueball superpotentials for SO(N) and USp(N),''
JHEP {\bf 0301}, 018 (2003)
[arXiv:hep-th/0211261].
}
\lref\KlemmCY{
A.~Klemm, K.~Landsteiner, C.~I.~Lazaroiu and I.~Runkel,
``Constructing gauge theory geometries from matrix models,''
JHEP {\bf 0305}, 066 (2003)
[arXiv:hep-th/0303032].
}


\lref\CachazoPR{F.~Cachazo and C.~Vafa,``N = 1 and N = 2
geometry from fluxes,'' arXiv:hep-th/0206017.
}

\lref\AIVW{M.~Aganagic, K.~Intriligator, C.~Vafa and
N.~P.~Warner,``The glueball
superpotential,'' arXiv:hep-th/0304271.
}
\lref\DijkgraafXK{R.~Dijkgraaf and C.~Vafa,``N = 1
supersymmetry, deconstruction, and bosonic gauge
theories,'' arXiv:hep-th/0302011.
}

\lref\Mehta{
M.~L.~Mehta,
``Random matrices,''
Academic Press (1991).
}

\lref\SinhaAP{
S.~Sinha and C.~Vafa,
``SO and Sp Chern-Simons at large N,''
arXiv:hep-th/0012136.
}
\lref\AhnVH{
C.~h.~Ahn, B.~Feng and Y.~Ookouchi,
``Phases of N = 1 SO(N(c)) gauge theories with flavors,''
arXiv:hep-th/0306068.
}

\lref\AshokBI{
S.~K.~Ashok, R.~Corrado, N.~Halmagyi, K.~D.~Kennaway and
C.~Romelsberger, ``Unoriented strings, loop equations, and N = 1
superpotentials from matrix models,''
Phys.\ Rev.\ D {\bf 67}, 086004 (2003)
[arXiv:hep-th/0211291].
}
\lref\IntriligatorFF{
K.~A.~Intriligator,
``New RG fixed points and duality in supersymmetric SP(N(c)) and SO(N(c)) gauge theories,''
Nucl.\ Phys.\ B {\bf 448}, 187 (1995)
[arXiv:hep-th/9505051].
}
\lref\DijkgraafFC{
R.~Dijkgraaf and C.~Vafa,
``Matrix models, topological strings, and supersymmetric gauge
theories,'' Nucl.\ Phys.\ B {\bf 644}, 3 (2002)
[arXiv:hep-th/0206255].
}
\lref\DijkgraafVW{
R.~Dijkgraaf and C.~Vafa,
``On geometry and matrix models,''
Nucl.\ Phys.\ B {\bf 644}, 21 (2002)
[arXiv:hep-th/0207106].
}
\lref\BalaTM{
V.~Balasubramanian, J.~de Boer, B.~Feng, Y.~H.~He, M.~x.~Huang, V.~Jejjala and A.~Naqvi,
``Multi-trace superpotentials vs. matrix models,''
Commun.\ Math.\ Phys.\  {\bf 242}, 361 (2003)
[arXiv:hep-th/0212082].
}
\lref\DijkgraafDH{
R.~Dijkgraaf and C.~Vafa,
``A perturbative window into non-perturbative physics,''
[arXiv:hep-th/0208048].
}
\lref\SeibergVS{
N.~Seiberg and E.~Witten,
``String theory and noncommutative geometry,''
JHEP {\bf 9909}, 032 (1999)
[arXiv:hep-th/9908142].
}
\lref\MinwallaPX{
S.~Minwalla, M.~Van Raamsdonk and N.~Seiberg,
``Noncommutative perturbative dynamics,''
JHEP {\bf 0002}, 020 (2000)
[arXiv:hep-th/9912072].
}
\lref\IntriligatorID{
K.~A.~Intriligator and N.~Seiberg,
``Duality, monopoles, dyons, confinement and oblique confinement in
supersymmetric SO(N(c)) gauge theories,''
Nucl.\ Phys.\ B {\bf 444}, 125 (1995)
[arXiv:hep-th/9503179].
}
\lref\WittenXY{
E.~Witten,
``Baryons and branes in anti de Sitter space,''
JHEP {\bf 9807}, 006 (1998)
[arXiv:hep-th/9805112].
}
\lref\GrossGK{
D.~J.~Gross and H.~Ooguri,
``Aspects of large N gauge theory dynamics as seen by string theory,''
Phys.\ Rev.\ D {\bf 58}, 106002 (1998)
[arXiv:hep-th/9805129].
}
\lref\DouglasNW{
M.~R.~Douglas and S.~H.~Shenker,
``Dynamics of SU(N) supersymmetric gauge theory,''
Nucl.\ Phys.\ B {\bf 447}, 271 (1995)
[arXiv:hep-th/9503163].
}

\lref\GubserFP{
S.~S.~Gubser and I.~R.~Klebanov,
``Baryons and domain walls in an N = 1 superconformal gauge theory,''
Phys.\ Rev.\ D {\bf 58}, 125025 (1998)
[arXiv:hep-th/9808075].
}
\lref\CsakiMX{
C.~Csaki, M.~Schmaltz, W.~Skiba and J.~Terning,
``Gauge theories with tensors from branes and orientifolds,''
Phys.\ Rev.\ D {\bf 57}, 7546 (1998)
[arXiv:hep-th/9801207].
}

\lref\CsakiEU{
C.~Csaki, W.~Skiba and M.~Schmaltz,
``Exact results and duality for Sp(2N) SUSY gauge theories with an  antisymmetric tensor,''
Nucl.\ Phys.\ B {\bf 487}, 128 (1997)[arXiv:hep-th/9607210].
}

\lref\LandsteinerXE{
K.~Landsteiner, C.~I.~Lazaroiu and R.~Tatar,
``Chiral field theories from conifolds,''
arXiv:hep-th/0310052.
}

\lref\AldayDK{
L.~F.~Alday and M.~Cirafici,
``Gravitational F-terms of SO/Sp gauge theories and anomalies,''
JHEP {\bf 0309}, 031 (2003)
[arXiv:hep-th/0306299].
}
\lref\AldayGB{
L.~F.~Alday and M.~Cirafici,
``Effective superpotentials via Konishi anomaly,''
JHEP {\bf 0305}, 041 (2003)
[arXiv:hep-th/0304119].
}

\lref\LandsteinerPH{
K.~Landsteiner and C.~I.~Lazaroiu,
``On Sp(0) factors and orientifolds,''
arXiv:hep-th/0310111.
}
\lref\LandsteinerRH{
K.~Landsteiner, C.~I.~Lazaroiu and R.~Tatar,
``(Anti)symmetric matter and superpotentials from IIB orientifolds,''
arXiv:hep-th/0306236.
}
\lref\BalasubramanianTV{ V.~Balasubramanian, B.~Feng, M.~x.~Huang and
A.~Naqvi, ``Phases of N = 1 supersymmetric gauge theories with
flavors,'' arXiv:hep-th/0303065.
}
\lref\MayrHH{
P.~Mayr,
``On supersymmetry breaking in string theory and its realization in
brane worlds,''
Nucl.\ Phys.\ B {\bf 593}, 99 (2001)
[arXiv:hep-th/0003198].
}
\lref\TaylorII{
T.~R.~Taylor and C.~Vafa,
``RR flux on Calabi-Yau and partial supersymmetry breaking,''
Phys.\ Lett.\ B {\bf 474}, 130 (2000)
[arXiv:hep-th/9912152].
}
\lref\GopakumarKI{
R.~Gopakumar and C.~Vafa,
``On the gauge theory/geometry correspondence,''
Adv.\ Theor.\ Math.\ Phys.\  {\bf 3}, 1415 (1999)
[arXiv:hep-th/9811131].
}
\lref\VafaWI{
C.~Vafa,
``Superstrings and topological strings at large N,''
J.\ Math.\ Phys.\  {\bf 42}, 2798 (2001)
[arXiv:hep-th/0008142].
}
\lref\CachazoJY{
F.~Cachazo, K.~A.~Intriligator and C.~Vafa,
``A large N duality via a geometric transition,''
Nucl.\ Phys.\ B {\bf 603}, 3 (2001)
[arXiv:hep-th/0103067].
}
\lref\CachazoZK{
F.~Cachazo, N.~Seiberg and E.~Witten,
``Phases of N = 1 supersymmetric gauge theories and matrices,''
JHEP {\bf 0302}, 042 (2003)
[arXiv:hep-th/0301006].
}
\lref\StromingerCZ{
A.~Strominger,
``Massless black holes and conifolds in string theory,''
Nucl.\ Phys.\ B {\bf 451}, 96 (1995)
[arXiv:hep-th/9504090].
}
\lref\dgkv{ R.~Dijkgraaf, S.~Gukov, V.~A.~Kazakov and C.~Vafa,
``Perturbative analysis of gauged matrix models,''
arXiv:hep-th/0210238.
}
%
\lref\CachazoZK{
F.~Cachazo, N.~Seiberg and E.~Witten,
``Phases of N = 1 supersymmetric gauge theories and matrices,''
JHEP {\bf 0302}, 042 (2003)
[arXiv:hep-th/0301006].
}
\lref\NaculichCZ{
S.~G.~Naculich, H.~J.~Schnitzer and N.~Wyllard,
``Cubic curves from matrix models and generalized Konishi anomalies,''
JHEP {\bf 0308}, 021 (2003)
[arXiv:hep-th/0303268].
}
\lref\NaculichKA{
S.~G.~Naculich, H.~J.~Schnitzer and N.~Wyllard,
``Matrix-model description of N = 2 gauge theories with
non-hyperelliptic Seiberg-Witten curves,''
arXiv:hep-th/0305263.
}
\lref\JanikNZ{
R.~A.~Janik and N.~A.~Obers,
``SO(N) superpotential, Seiberg-Witten curves and loop equations,''
Phys.\ Lett.\ B {\bf 553}, 309 (2003)
[arXiv:hep-th/0212069].
}
\lref\AhnCQ{
C.~h.~Ahn and Y.~Ookouchi,
``Phases of N = 1 supersymmetric SO / Sp gauge theories via matrix
model,'' JHEP {\bf 0303}, 010 (2003)
[arXiv:hep-th/0302150].
}
\lref\AhnUI{
C.~h.~Ahn, B.~Feng and Y.~Ookouchi,
``Phases of N = 1 USp(2N(c)) gauge theories with flavors,''
arXiv:hep-th/0307190.
}
\lref\FujiVV{
H.~Fuji and Y.~Ookouchi,
``Confining phase superpotentials for SO/Sp gauge theories via
geometric transition,''
JHEP {\bf 0302}, 028 (2003)
[arXiv:hep-th/0205301].
}
\lref\EdelsteinMW{
J.~D.~Edelstein, K.~Oh and R.~Tatar,
``Orientifold, geometric transition and large N duality for SO/Sp gauge
theories,''
JHEP {\bf 0105}, 009 (2001)
[arXiv:hep-th/0104037].
}
\lref\ChoBI{
P.~L.~Cho and P.~Kraus,
``Symplectic SUSY gauge theories with antisymmetric matter,''
Phys.\ Rev.\ D {\bf 54}, 7640 (1996)
[arXiv:hep-th/9607200].
}
\lref\KrausJV{
P.~Kraus, A.~V.~Ryzhov and M.~Shigemori,
``Loop equations, matrix models, and N = 1 supersymmetric gauge
theories,'' JHEP {\bf 0305}, 059 (2003)
[arXiv:hep-th/0304138].
}
\lref\KrausJF{
P.~Kraus and M.~Shigemori,
``On the matter of the Dijkgraaf-Vafa conjecture,''
JHEP {\bf 0304}, 052 (2003)
[arXiv:hep-th/0303104].
}
\lref\CachazoKX{
F.~Cachazo,
``Notes on supersymmetric Sp(N) theories with an antisymmetric tensor,''
arXiv:hep-th/0307063.
}
\lref\FengGB{
B.~Feng,
``Geometric dual and matrix theory for SO/Sp gauge theories,''
Nucl.\ Phys.\ B {\bf 661}, 113 (2003)
[arXiv:hep-th/0212010].
}
\lref\cdsw{F.~Cachazo, M.~R.~Douglas, N.~Seiberg and
E.~Witten,``Chiral rings and anomalies in supersymmetric gauge
theory,''JHEP {\bf 0212}, 071 (2002)[arXiv:hep-th/0211170].
}

\lref\dglvz{R.~Dijkgraaf, M.~T.~Grisaru, C.~S.~Lam, C.~Vafa
and D.~Zanon,``Perturbative computation of glueball
superpotentials,'' arXiv:hep-th/0211017.
}

\lref\SeibergRS{
N.~Seiberg and E.~Witten,
``Electric - magnetic duality, monopole condensation, and confinement in
N=2 supersymmetric Yang-Mills theory,''
Nucl.\ Phys.\ B {\bf 426}, 19 (1994)
[Erratum-ibid.\ B {\bf 430}, 485 (1994)]
[arXiv:hep-th/9407087].
}
\lref\SeibergAJ{
N.~Seiberg and E.~Witten,
``Monopoles, duality and chiral symmetry breaking in N=2 supersymmetric
QCD,''
Nucl.\ Phys.\ B {\bf 431}, 484 (1994)
[arXiv:hep-th/9408099].
}
\lref\CsakiVV{
C.~Csaki and H.~Murayama,
``Instantons in partially broken gauge groups,''
Nucl.\ Phys.\ B {\bf 532}, 498 (1998)
[arXiv:hep-th/9804061].
}
\lref\dohol{ N.~Dorey, T.~J.~Hollowood and S.~Prem Kumar, ``An
exact elliptic superpotential for N = 1* deformations of finite  N
= 2 gauge theories,'' Nucl.\ Phys.\ B {\bf 624}, 95 (2002)
[arXiv:hep-th/0108221].
,
}
\lref\Doreyb{ N.~Dorey, T.~J.~Hollowood, S.~P.~Kumar and
A.~Sinkovics, ``Massive vacua of N = 1* and S-duality from matrix
models,'' JHEP {\bf 0211}, 040 (2002) [arXiv:hep-th/0209099].
}
\lref\DoreyTJ{ N.~Dorey, T.~J.~Hollowood, S.~Prem Kumar and A.~Sinkovics, ``Exact superpotentials from matrix models,'' JHEP {\bf
0211}, 039 (2002) [arXiv:hep-th/0209089].
}

\Title{
  \vbox{\baselineskip12pt \hbox{hep-th/0311181}
  \hbox{BRX TH-525}
  \hbox{HUTP-03/A073}
  \hbox{UCLA-03-TEP-29}
  \hbox{UCSD-PTH-03-14}
  \vskip-.5in}
}{\vbox{
  \centerline{On Low Rank Classical Groups in}
  \centerline{ String Theory, Gauge Theory and Matrix Models}
}}

\centerline{Ken Intriligator$^1$, Per Kraus$^2$, Anton V. Ryzhov$^3$,
Masaki Shigemori$^2$ and Cumrun Vafa$^4$}
\bigskip\medskip
\centerline{$^1$ \it Department of Physics, University of California, San Diego, La Jolla, CA 92093-0354, USA}
\vskip-.05in
\centerline{$^2$ \it Department of Physics and Astronomy, UCLA, Los Angeles, CA 90095-1547, USA}
\vskip-.05in
\centerline{$^3$ \it Department of Physics, Brandeis University, Waltham, MA 02454, USA}
\vskip-.05in
\centerline{$^4$ \it Jefferson Physical Laboratory, Harvard University, Cambridge, MA 02138, USA}
\medskip
\medskip
\medskip
\medskip
\medskip
\medskip
\baselineskip14pt
\noindent

We consider ${\cal N}=1$ supersymmetric $U(N)$, $SO(N)$, and $Sp(N)$
gauge theories, with two-index tensor matter and added tree-level
superpotential, for general breaking patterns of the gauge group.  By
considering the string theory realization and geometric transitions, we
clarify when glueball superfields should be included and extremized, or
rather set to zero; this issue arises for unbroken group factors of low
rank.  The string theory results, which are equivalent to those of the
matrix model, refer to a particular UV completion of the gauge theory,
which could differ from conventional gauge theory results by residual
instanton effects.  Often, however, these effects exhibit miraculous
cancellations, and the string theory or matrix model results end up
agreeing with standard gauge theory.
In particular, these string theory considerations explain and remove
some apparent discrepancies between gauge theories and matrix models in
the literature.

\Date{November, 2003}
\baselineskip14pt
\newsec{Introduction}

Large $N$ topological string duality \GopakumarKI\ embedded in
superstrings \refs{\VafaWI, \CachazoJY} has led to a new perspective on
${\cal N}=1$ supersymmetric gauge theories: that the exact effective
superpotential can be efficiently computed by including glueball
fields. For example, in a theory with gauge group $G$, with tree-level
superpotential leading to a breaking pattern
\eqn\higgsbp{G(N)\rightarrow \prod _{i=1}^K G_i(N_i),}
the dynamics is
efficiently encoded in a superpotential $W_{\rm eff}(S_1, \dots S_K; g_j,
\Lambda )$ ($g_j$ are the parameters in $W_{\rm tree}$ and $\Lambda$ is the
dynamical scale).  Further, string theory implies \CachazoJY
\eqn\weffgen{
  W_{\rm eff}(S_i; g_j, \Lambda)=
  \sum _{i=1}^K
  \left(
    h_i{\partial {\cal F}(S_i) \over \partial S_i}-2\pi i \tau _i S_i
  \right),
}
 with $h_i$ and $\tau _i$ the fluxes through $A_i$ and $B_i$
three-cycles in the geometry, as will be reviewed in sect.\ 3.  The
prepotential ${\cal F}(S_i)$ in \weffgen\ is computable in terms of
geometric period integrals, which yields \CachazoJY\
\eqn\prepotgen{
  {\partial {\cal F}(S_i)\over \partial S_i}
  =
  S_i \left(\log \left({\Lambda _i^3\over S_i}\right)+1\right)
  +{\partial \over \partial S_i}
  \sum_{i_1, \dots , i_K\geq 0}c_{i_1\dots i_K}S_1^{i_1}\cdots S_K^{i_k},
}
with coefficients $c_{i_1\dots i_K}$ depending on the $g_j$ (but not on
the gauge theory scale $\Lambda$).  In \DijkgraafFC\ it was shown how
planar diagrams of an associated matrix model can also be used to
compute \weffgen\ and \prepotgen.  Based on the stringy examples, this
was generalized in \DijkgraafDH\ to a more general principle to gain
non-perturbative information about the strong coupling dynamics of gauge
theories, by extremizing the perturbatively computed glueball
superpotential.

There are two aspects to the above statements: first that the glueball
fields $S_i$ are the `right' variables to describe the IR physics, and
second that perturbative gauge theory techniques suffice to compute the
glueball superpotential.  The latter statement has now been proven in
two different approaches for low powers of the glueball fields $S_i$ in
\prepotgen\ \refs{\dglvz, \cdsw}.  For powers of the glueball fields
$S_i$ larger than the dual Coxeter number of the group, an ambiguity
sets in for the glueball computation of the coefficients $c_{i_1\dots
i_K}$ in both of these approaches.  The matrix model provides a natural
prescription for how to resolve this ambiguity, essentially by
continuing from large $N_i$.  It was argued in \refs{\DijkgraafXK,
\AIVW}\ that the string geometry / matrix model result (since the string
geometry and matrix model results are identical, we refer to them
synonymously) has the following meaning: it computes the $F$-terms for
different supersymmetric gauge theories, which can be expressed in terms
of $G(N+k|k)$ supergroups.  The $W_{\rm eff}(S_i)$ is independent of
$k$, and the above ambiguity can be eliminated by taking $k$, and hence
the dual Coxeter number, arbitrarily large.  The $G(N)$ theory of
interest is obtained from the $G(N+k|k)$ theory by Higgsing; but there
can be residual instanton contributions to $W_{\rm eff}$ \AIVW, which
can lead to apparent discrepancies between the matrix model and gauge
theory results.  We will somewhat clarify here when such residual
instanton effects do, or do not, lead to discrepancies with standard
gauge theory results.

There is another, more non-trivial assumption in \DijkgraafDH : the
statement that the glueball fields $S_i$ are the `right' variables in
the IR.  This assumption was motivated from the string dualities
\refs{\GopakumarKI - \CachazoJY}, where geometric transition provide the
explanation of why the glueball fields are the natural IR variables:
heuristically, $\ev{S_i}$ corresponds to confinement.  However, this is
not quite correct: it also applies to abelian theories, as had been
noted in \CachazoJY . So the deep explanation of why we should choose
certain dynamical $S$ variables remains mysterious.

In this paper, we will uncover the precise prescription for the correct
choice of IR variables.  This will be done from the string theory
perspective, by arguing in which cases there is a geometric transition
in string theory.  For the general breaking pattern \higgsbp, our
prescription for treating the glueball field $S_i$, corresponding to the
factor $G_i$ in \higgsbp, is as follows: {\it If $h(G_i)>0$ we include
$S_i$ and extremize $W_{\rm eff}(S_i)$ with respect to it.  On the other
hand, if $h(G_i)\leq 0$ we do not include or extremize $S_i$, instead we
just set $S_i\rightarrow 0$.}  Here we define the generalized
dual-Coxeter numbers\foot{Our convention is such that $Sp(N)\subset
SU(2N)$, and hence $Sp(1)\cong SU(2)$.}
\eqn\wlowuiv{\eqalign{
  h(U(N))  &= N,\cr
  h(Sp(N)) &= N+1,\cr
  h(SO(N)) &= N-2,
}}
which are generalized in that \wlowuiv\ applies for all $N\geq 0$.  In
particular $h(U(1))= h(Sp(0))=1$, so when some $G_i$ factor in \higgsbp\
is $U(1)$ or $Sp(0)$, our prescription is to include the corresponding
$S_i$ and extremize with respect to it. On the other hand, $h(U(0))=0$
and $h(SO(2))=0$, so when some $G_i$ factor in \higgsbp\ is $U(0)$ or
$SO(2)$, our prescription is to just set the corresponding $S_i=0$ from
the outset.  (Note that $U(1)$ and $SO(2)$ are treated differently
here.)

This investigation was motivated by trying to understand the
discrepancies found in \KrausJF\ for $Sp(N)$ theory with antisymmetric
tensor matter, where the superpotentials from the matrix model and gauge
theory were found to differ at order $h$ in perturbation theory and
beyond.  The analysis considered the trivial breaking pattern
$Sp(N)\rightarrow Sp(N)$ and a single glueball was introduced
corresponding to the single unbroken gauge group factor.
In \AIVW, various gauge theories including this example were studied,
and an explanation for the discrepancies was proposed in terms of
the conjecture, mentioned above,  that the string theory / matrix model 
actually computes the superpotential of the large $k$ $G(N+k|k)$ supergroup 
theories, rather than the ordinary $G(N)$ theory.  
In this context, the trivial breaking pattern considered in \KrausJF\
should be understood as $Sp(N)\to Sp(N)\times Sp(0)$, which is completed
to $Sp(N+k|k)\to Sp(N+k_1|k_1)\times Sp(k_2|k_2)$.  In particular,
$Sp(0)$ factors, while trivial in standard gauge theory, are non-trivial
in the string theory geometry / matrix model context:
there can be a residual instanton contribution to the superpotential
when one Higgses $Sp(k_2|k_2)$ down to $Sp(0|0)=Sp(0)$, as explicitly
seen in \AIVW\ for the case of breaking $Sp(0)\rightarrow Sp(0)$ with
quadratic $W_{\rm tree}$\foot{ It was suggested in the original version
of \AIVW\ that such $Sp(0)$ residual instanton contributions could also
play a role for the case of cubic and higher order $W_{\rm tree}$ (where
they had not yet been fully computed) and could explain the apparent
matrix model vs.\ standard gauge theory discrepancies found in \KrausJF.
As we will discuss, we now know that this last speculation was not
correct.  The corrected proposal of \AIVW\ is still that the matrix
model computes the superpotential of the $G(N+k|k)$ theory, but where
the matrix model side of the computation should be corrected, as we
discuss in this paper, to include glueball fields for the $Sp(0)$
factors.}.  Related aspects of ``$Sp(0)$'' being non-trivial in the
string / matrix model context were subsequently discussed in
\refs{\CachazoKX, \AhnUI, \LandsteinerPH}.

However, it turns out that one also needs to modify the matrix model
side of the computation to take into account the $Sp(0)$ factors. 
This was found by Cachazo \CachazoKX,
who showed that the loop equations determining $T(z)\equiv
\Tr({1\over z-\Phi})$ and $R(z)\equiv -{1\over 32\pi
^2}\Tr({W_\alpha W^\alpha \over z-\Phi})$ for the $Sp(N)$ theory with
antisymmetric tensor matter \refs{\AldayGB, \KrausJV}\ could be related
to those of a $U(2N+2K)$ gauge theory with adjoint matter, with $Sp(N)
\rightarrow Sp(N)\times Sp(0)^{K-1}$ mapped to $U(2N+2K)\rightarrow
U(2N+2) \times U(2)^{K-1}$.
It was thus shown in \CachazoKX\ that vanishing period of $T(z)dz$
through a given cut, corresponding to an $Sp(0)$ factor, does not imply
that the cut closes up on shell (aspects of the periods in this theory
were also discussed in \MatoneBX).
This fits with our above prescription
that the $Sp(0)$ glueballs should be included and extremized in the
string theory / matrix model picture, as would be done for $U(2)$,
rather than set to zero, as was originally done in \KrausJF.  We stress
that we are not yet even discussing whether or not the string theory /
matrix model result agrees with standard gauge theory.  Irrespective of
any comparison with standard gauge theory, the prescription to obtain
the actual string theory / matrix model result is as described above
\wlowuiv.  Having obtained that result, we can now discuss comparisons
with standard gauge theory results.  As seen in \CachazoKX, by solving
the $U(2N+2K)$ loop equations for the present case, this corrected
matrix model result now agrees perfectly with standard gauge theory!
This will be discussed further here, with all glueball fields $S_i$
included.

This agreement, between the matrix model result and standard gauge
theory, is in a sense surprising for this particular theory, in light of
the $Sp(k|k)$ description of \AIVW\ for the unbroken $Sp(0)$ factors,
with the resulting residual instanton contributions to the
superpotential.  As we will explain later in this paper, the agreement
here between matrix models and standard gauge theory is thanks to a
remarkable cancellation of the residual instanton effect terms, which
could have spoiled the agreement.  The cancellation occurs upon summing
over the $i$ in \weffgen\ from $i=1\dots K$.

There are similar remarkable cancellations of the ``residual instanton
contributions'' to the superpotential in many other examples, which we
will also discuss.  In fact, in all cases that we know of, the only
cases where the residual instantons do not cancel is when the gauge
theory clearly has some ambiguity, requiring a choice of how to define
the theory in the UV; string theory / matrix model gives a particular
such choice.  Examples of such cases is when the LHS of \higgsbp\ is
itself $U(1)$ or $Sp(0)$ super Yang-Mills, as discussed in \AIVW.  Other
examples where the residual instanton contributions do not cancel, is
when the superpotential is of high enough order such that not all
operators appearing in it are independent, e.g.\ terms like $\Tr\, \Phi
^n$, for a $U(N)$ adjoint $\Phi$, when $n>N$.  In standard gauge theory,
there are then potential ambiguities involved in reducing such composite
operators to the independent operators, since classical operator
identities can receive quantum corrections.  The residual instanton
contributions, which do not cancel generally in these cases, imply
specific quantum relations for these operators, corresponding to the
specific UV completion.  See \refs{\DoreyTJ, \BalaTM} for related
issues.

The organization of this paper is as follows: In section 2 we summarize
the gauge theories under consideration. In section 3 we review the type
IIB string theory construction of these gauge theories.  We also discuss
maps of the exact superpotentials of $Sp$ and $SO$ theories to those of
$U$ theories, generalizing observations of \refs{\ItaKX, \AshokBI,
\JanikNZ, \CachazoKX, \LandsteinerPH}.  In section 4 we explain, from
the string theory perspective in which cases we have a geometric
transition. In section 5 we consider examples, where the glueball fields
$S_i$ of all group factors are correctly accounted for on the matrix
model side.  The results thereby obtained via matrix models are found to
agree with those of standard gauge theory.  In many of these examples,
this agreement relies on a remarkable interplay of different residual
instanton contributions, which sometimes fully cancel.  Residual
instantons are discussed further in sect.\ 6, with examples illustrating
cases where they do, or do not, cancel.  In appendix A, a proof of a
general relation between the $S^2$ and $RP^2$ contributions to the
matrix model free energy is given, and also the matrix model computation
of superpotential is presented.  In appendix B the gauge theory
computation of the superpotential is discussed.

\newsec{The gauge theory examples}

The specific examples of ${\cal N}=1$ supersymmetric gauge theories
which we consider, with breaking patterns as in \higgsbp, are as
follows:
\eqn\higgscases{
\matrix{&U(N)&\hbox{with adjoint}\ \Phi\qquad &
U(N)\rightarrow \prod _{i=1}^KU(N_i),\qquad&
\cr
&SO(N)&\hbox{with adjoint}\ \Phi: \qquad & SO(N)\rightarrow SO(N_0)\times \prod _{i=1}^{K}U(N_i), \qquad\cr
&Sp(N) &\hbox{with adjoint}\ \Phi: \qquad &Sp(N)\rightarrow Sp(N_0)\times \prod _{i=1}^{K}U(N_i),\qquad\cr
&SO(N)&\hbox{with symmetric}\ S:\qquad & SO(N)\rightarrow \prod _{i=1}^K SO(N_i),\qquad
\cr
&Sp(N)&\hbox{with antisymmetric}\ A:\qquad & Sp(N)\rightarrow \prod _{i=1}^K Sp(N_i),\qquad \cr
&U(N)&\hbox{with}\ \Phi+S+\widetilde S:\qquad &U(N)\rightarrow SO(N_0)\times
\prod _{i=1}^{K} U(N_i),\qquad\cr
&U(N)&\hbox{with}\ \Phi+A+\widetilde A:\qquad &U(N)\rightarrow Sp(N_0)\times
\prod _{i=1}^{K} U(N_i).\qquad\cr
}}
For $U(N)$ with adjoint $\Phi$, the tree-level superpotential is taken to be 
\eqn\ba{\eqalign{
  W_{\rm tree} &= \Tr[W(\Phi)],\qquad
  W(x) = \sum_{j=1}^{K+1} {g_j\over j}x^j,
}}
with $K$ potential wells.    In the classical vacua, with breaking pattern as in 
\higgscases, $\Phi$ has $N_i$ eigenvalues equal to the root
$a_i$ of
\eqn\Wprimei{
 W'(x)=\sum _{j=1}^{K+1}g_{j}x^{j-1}\equiv g_{K+1}\prod_{i=1}^K(x-a_i),
}
with $\sum _{i=1}^K N_i=N$.  For $SO(N)$ with symmetric tensor or
$Sp(N)$ with antisymmetric tensor we take
\eqn\bsospx{
  W_{\rm tree}=\half \Tr W(S), 
  \qquad \hbox{or} \qquad
  W_{\rm tree}=\half \Tr W(A),
}
respectively, where $W(x)$ is as in \ba, the factor of $\half$ is for convenience,
because the eigenvalues of $S$ or $A$ appear in pairs, 
and the indices are contracted with $\delta ^a_b$ for
$SO(N)$ or $J^a_b$ for $Sp(N)$.  
For $SO(N)$ and $Sp(N)$ with adjoint matter, the tree-level superpotential is 
\eqn\basp{\eqalign{
  W_{\rm tree} &= \half \Tr[W(\Phi)],\qquad
  W(x) = \sum_{j=1}^{K+1} {g_{2j}\over 2j}x^{2j},
}}
since all Casimirs of the adjoint $\Phi$ are even, and the $\half$ is
again for convenience because the eigenvalues appear in pairs.  $\Phi$'s
eigenvalues sit at the zeros of
\eqn\Wprimeisp{
  W'(x)=\sum _{j=1}^{K+1}g_{2j}x^{2j-1}\equiv
  g_{2K+2}x\prod _{i=1}^{K}(x^2-a_i^2).
}
The breaking pattern in \higgscases\ has $N_0$ eigenvalues of $\Phi$
equal to zero, and $N_i$ pairs at $\pm a_i$, so $N=N_0+\sum
_{i=1}^{K}2N_i$ for $SO(N)\rightarrow SO(N_0)\times \prod
_{i=1}^{K}U(N_i)$, and $N=N_0+\sum _{i=1}^{K}N_i$ for
$Sp(N)\rightarrow Sp(N_0)\times \prod_{i=1}^{K}U(N_i)$ (with the
convention $Sp(1)\cong SU(2)$).

The next to last example in \higgscases\ is the ${\cal N}=2$ $U(N)$
theory with a matter hypermultiplet in the two-index symmetric tensor
representation, breaking ${\cal N}=2$ to ${\cal N}=1$ by a
superpotential as in \ba:
\eqn\baa{W=\sum _{j=1}^{K+1}{g_j\over j}\Tr \Phi ^j +\sqrt{2}\Tr
\widetilde S \Phi S.}
In addition to the possibility of $\Phi$'s eigenvalues sitting in any of
the $K$ critical points $W'(x)$ analogous to \Wprimei, there is a vacuum
where $N_0$ eigenvalues sits at $\phi =0$, with $\ev{S\widetilde S}\neq
0$, breaking $U(N_0)\rightarrow SO(N_0)$.  The last example in
\higgscases\ is the similar theory where the ${\cal N}=2$ hypermultiplet
is instead in the antisymmetric tensor representation $A$, rather than
the symmetric tensor $S$.  These last two classes of examples were
considered in \refs{\KlemmCY,\NaculichCZ, \NaculichKA}.

In all of these theories, the low energy superpotential is of the
general form
\eqn\wlowg{
W_{\rm low}(g_j, \Lambda)=W_{\rm cl}(g_j)+W_{\rm gc}(\Lambda _i)+W_H(g_j,
\Lambda).}
$W_{\rm cl}(g_j)$ is the classical contribution (evaluating $W_{\rm
tree}$ in the appropriate minima).  $W_{\rm gc}(\Lambda _i)$ is the
gaugino condensation contribution in the unbroken gauge groups of
\higgsbp,
\eqn\wgcisi{W_{\rm gc}(\Lambda _j)=\sum _{i=1}^K h_ie^{2\pi i
n_i/h_i}\Lambda _i^3,}
where $h_i=C_2(G_i)$ is the dual Coxeter number of the group $G_i$ in
\higgsbp, with the phase factors associated with the ${\bf
Z}_{2h_i}\rightarrow {\bf Z}_2$ chiral symmetry breaking of the
low-energy $G_i$ gaugino condensation.  The scales $\Lambda _i$ are
related to $\Lambda$ by threshold matching for the fields which got a
mass from $W_{\rm tree}$ and the breaking \higgsbp; some examples, with
breaking patterns as in \higgscases, are as follows.  For $U(N)$ with
adjoint $\Phi$:
\eqn\lamin{\Lambda _i^{3N_i}=\Lambda ^{2N} W''(a_i)^{N_i}
\prod _{j\neq i}m_{W_{ij}}^{-2N_j}
=g_{K+1}^{N_i} \Lambda ^{2N}
\prod _{j\neq i}(a_j-a_i)^{N_i-2N_j}.}
For $SO(N)$ with adjoint, breaking $SO(N)\rightarrow SO(N_0)\times \prod
_{i=1}^{K}U(N_i)$,
\eqn\lamsoa{\eqalign{\Lambda _0^{3(N_0-2)}&=g_{2K+2}^{N_0-2}\Lambda ^{2(N-2)}
\prod _{i=1}^{K}a_i^{2(N_0-2)-4N_i}\cr
\Lambda _i^{3N_i}&=2^{-N_i}g_{2K+2}^{N_i}\Lambda ^{2(N-2)}a_i^{-2(N_0-2)}\prod _{ j\neq i}^{K}
(a_i^2-a_j^2)^{N_i-2N_j}.}}
For $Sp(N)$ with adjoint
\eqn\lamspa{\eqalign{\Lambda _0^{3(N_0+1)}&=g_{2K+2}^{N_0+1}\Lambda ^{2(N+1)}\prod _{i=1}^{K}
a_i^{2(N_0+1)-2N_i}\cr
\Lambda _i^{3N_i}&=2^{-N_i}g_{2K+2}^{N_i}\Lambda ^{4(N+1)}a_i^{-4(N_0+1)}
\prod _{j\neq i}^{K}(a_i^2-a_j^2)
^{N_i-2N_j};}}
the $U(N_i)$ has index of embedding 2, which is why the $U(N_i)$
one-instanton factor is related to the $Sp(N)$ two-instanton factor in
\lamspa.  For $SO(N)$ with symmetric tensor,
\eqn\lamsos{\Lambda _i ^{3(N_i-2)}=g_{K+1}^{N_i+2}\Lambda ^{2N-8}\prod _{j\neq i=1}^K
(a_i-a_j)^{N_i+2-2N_j}.}
For $Sp(N)$ with antisymmetric $A$:
\eqn\laminsp{\Lambda _i^{3(N_i+1)}=g_{K+1}^{N_i-1}\Lambda ^{2N+4}\prod _{j\neq
i}^K(a_i-a_j)^{N_i-1-2N_j}.}

Finally, the term $W_H(g_j, \Lambda)$ in \wlowg\ are additional
non-perturbative contributions, which can be regarded as coming from the
massive, broken parts of the gauge group.

In the description with the glueballs $S_i$ integrated in, as in
\weffgen, the gaugino condensation contribution comes from the first
term in \prepotgen:
\eqn\wgca{W_{\rm gc}(S_i, \Lambda)=\sum _{i=1}^K h_i S_i \left(\log
\left({\Lambda _i^3 \over S_i}\right)+1\right),}
and $W_H(g_i, \Lambda)$ comes from the last terms in \prepotgen, upon
integrating out the $S_i$.  When the minima $a_i$ of the superpotential
are widely separated, the contributions $W_H$ from these last terms are
subleading as compared with $W_{\rm gc}$.  As in \weffgen, the full
glueball superpotential \weffgen\ can be computed via the string theory
geometric transition, in terms of certain period integrals
\refs{\VafaWI, \CachazoJY}, as will be reviewed in the next section, or
via the matrix models.  In that context, the term $W_{\rm gc}(S_i,
\Lambda _i)$ comes from the integration measure (as is also natural in
field theory, since it incorporates the $U(1)_R$ anomaly) and $W_H(S_i,
g_j)$ can be computed perturbatively \refs{\DijkgraafFC, \DijkgraafDH}.
The perturbative computation of $W_H(S_i, g_j)$ can also be understood
directly in the gauge theory \dglvz, up to ambiguities in terms $S^n$
with $n>h=C_2(G)$.  The string theory/ matrix model constructions yield
a specific way of resolving these ambiguities, which correspond to a
particular UV completion of the gauge theory \refs{\DijkgraafXK, \AIVW}.

As discussed in the introduction, our present interest will be in
analyzing this circle of ideas when some of the gauge group factors in
\higgsbp\ are of low rank, or would naively appear to be trivial, e.g.\
$U(1)$, $U(0)$, $SO(2)$, $SO(0)$, and $Sp(0)$.

\newsec{Geometric transition of U(N) and SO/Sp(N) theories}

In this section we briefly review the type IIB geometric engineering of
relevant $U(N)$ and $SO/Sp(N)$ theories, and their geometric transition.

%
\bigskip
\subsec{$U(N)$ with adjoint and $W_{\rm tree}=\Tr \sum_{j=1}^{K+1}{g_i\over j}\Phi ^j$}

The Calabi--Yau geometry relevant to this theory \CachazoJY\ is the
non-compact $A_1$ fibration
\eqn\bb{\eqalign{
  W'(x)^2+y^2+u^2+v^2=0.
}}
This fibration has $K$ conifold singularities at the critical points of
$W(x)$, i.e.\ at $W'(x)=0$.  Near each of the singularities, the
geometry \bb\ is the same as the usual conifold $x'{}^2+y^2+u^2+v^2=0$,
which is topologically a cone with base $S^2\times S^3$.

The singularities can be resolved by blowing up a 2-sphere $S^2=\Pone$
at each singularity.  We can realize the $U(N)$ gauge theory with
adjoint matter and superpotential \ba\ in type IIB superstring theory
compactified on this resolved geometry, with $N$ D5-branes partially
wrapping the $K$ $\Pone$'s.  The gauge theory degrees of freedom
correspond to the open strings living on these D5-branes.  The classical
supersymmetric vacuum is obtained by distributing the $N_i$ D5-branes
over the $i$-th 2-sphere $\Pone_i$ with $i=1,\cdots,K$.  The
corresponding breaking pattern of the gauge group is as in \higgscases:
$U(N)\to \prod _{i=1}^KU(N_i)$.

At low energy, the gauge theory confines (when $N_i>1$), each $U(N_i)$
factor developing nonzero vev of the glueball superfield $S_i$.  In
string theory this is described by the geometric transition
\refs{\GopakumarKI , \VafaWI , \CachazoJY}\ in which the resolved
conifold geometry with $\Pone$'s wrapped by D5-branes is replaced by a
deformed conifold geometry
\eqn\bc{\eqalign{
  W'(x)^2+f_{K-1}(x)+y^2+u^2+v^2=0,
}}
where $f_{K-1}(x)$ is a polynomial of degree $K-1$ in $x$ and
parametrizes the deformation.  After the geometric transition, each
2-sphere $\Pone_i$ wrapped by $N_i$ D5-branes is replaced by a 3-sphere
$A_i$ with 3-form RR flux through it:
\eqn\bd{\eqalign{
  \oint_{A_i} H = N_i,
}}
where $H = H_{RR} + \tau H_{NS}$ and $\tau=C^{(0)}+ie^{-\Phi}$ is the
complexified coupling constant of type IIB theory.
We define the periods of the Calabi--Yau geometry \bc\ by
\eqn\be{\eqalign{
  S_i \equiv  {1\over 2\pi i}\oint_{A_i} \Omega,
  \qquad
  \Pi_i \equiv \int_{B_i}^{\Lambda_b} \Omega,
}}
where $\Omega$ is the holomorphic 3-form, $B_i$ is the noncompact
3-cycle dual to the 3-cycle $A_i$, and $\Lambda_b$ is a cutoff needed to
regulate the divergent $B_i$ integrals. The IR cutoff $\Lambda_b$ is to
be identified with the UV cutoff of the 4d gauge theory.  The set of
variables $S_i$ measure the size of the blown up 3-spheres, and can be
used to parametrize the deformation in place of the $K$ coefficients of
the polynomial $f_{K-1}(x)$.

The dual theory after the geometric transition is described by a 4d,
$\CN=1$ $U(1)^K$ gauge theory, with $K$ $U(1)$ vector superfields $V_i$
and $K$ chiral superfields $S_i$.  If not for the fluxes, this theory
would be ${\cal N}=2$ supersymmetric $U(1)^K$, with $(V_i, S_i)$ the
${\cal N}=2$ vector super-multiplets, $S_i$ being the Coulomb branch
moduli.  This low-energy $U(1)^K$ theory has a non-trivial prepotential
${\cal F}(S_i)$ and the dual periods $\Pi _i$ in \be\ can be written as
\eqn\Piper{\Pi _i (S)={\partial {\cal F}\over \partial S_i}.}  Without
the fluxes, this prepotential can be understood as coming {}from
integrating out D3 branes which wrap the $A_i$ cycles, and are charged
under the low-energy $U(1)$'s \StromingerCZ.

The effect of the added fluxes is to break ${\cal N}=2$ supersymmetry to
${\cal N}=1$ by the added superpotential \refs{\TaylorII , \MayrHH}
\eqn\bg{\eqalign{
  W_{\rm flux}
  =
  \int H\wedge \Omega
  =
  \sum_i
  \left(
    \oint_{A_i} H \int_{B_i}^{\Lambda_b} \Omega
    -\int_{B_i}^{\Lambda_b} H \oint_{A_i} \Omega
  \right).
}}
In the present $U(N)$ case, \bd\ and \be\ gives
\eqn\bga{\eqalign{
  W_{\rm flux}
  =
  \sum_{i=1}^K (N_i \Pi_i  - 2\pi i \alpha S_i),
}}
where
\eqn\bh{\eqalign{
  \oint_{B_i}^{\Lambda_b} H = \alpha
}}
is the 3-form NS flux through the 3-cycle $B_i$ and identified
with the bare coupling constant of the gauge theory by
\eqn\bha{\eqalign{
  2\pi i \alpha  =  {8\pi^2 \over g_b^2}=V.
}}
where $V$ is the complexified volume of the ${\bf P}^1$'s.  The ${\cal
N}=1$ $U(1)^K$ vector multiplets $V_i$ remain massless, but the $S_i$
now have a superpotential, which fixes them to sit at discrete vacuum
expectation values, where they are massive.  The fields $S_i$ are
identified with the glueballs on the gauge theory side.

The superpotential $W_{\rm flux}(S_i)$ is the full, exact, effective
superpotential in \weffgen.  As can be verified by explicit calculations
\refs{\VafaWI ,\CachazoJY}, the leading contribution to \bga\ is always
of the form
\eqn\bi{\eqalign{
  W_{\rm flux} \sim \sum_{i=1}^K [N_i S_i(1-\ln(S_i/\Lambda_i^3)) - 2
\pi
i \alpha S_i],
}}
where $\Lambda_i$ is related to the scale $\Lambda_b$ via precisely the
relation \lamin.  This leading term \bi\ is the gaugino condensation
part of the superpotential, as in \prepotgen.

%
\bigskip
\subsec{$SO$ and $Sp$ theories}

The string theory construction of $SO/Sp(N)$ theories can be obtained
from the above $U(N)$ construction, by orientifolding the geometry
before and after the geometric transition by a certain $\Ztwo$ action.
The geometric construction of $SO/Sp(N)$ theory with adjoint was
discussed in \refs{\SinhaAP ,\EdelsteinMW ,\AshokBI ,\LandsteinerRH},
and in that case the invariance of the geometry \bc\ under the $\Ztwo$
action requires that the polynomial $W(x)$ be even.  The geometric
construction of $SO/Sp(N)$ theory with symmetric/antisymmetric tensor
was studied in \refs{\CsakiMX , \LandsteinerXE, \LandsteinerPH}.

In the classical vacuum of the ``parent'' $U(2N)$ theory, the gauge
group is broken into a product of $U(N_i)$ groups.  When a $U(N_i)$
factor is identified with another $U(N_i)$ by the $\Ztwo$ orientifold
action, they lead to a single $U(N_i)$ factor.  When a $U(N_i)$ factor
is mapped to itself by the $\Ztwo$ orientifold action, it becomes an
$SO(N_i)$ or $Sp(\half N_i)$, depending on the charge of the orientifold
hyperplane.  As a result, the classical vacuum of the ``daughter''
$SO/Sp(N)$ theory has gauge group broken as in \higgscases, depending on
whether the theory is $SO$ or $Sp$ with adjoint, or $SO$ with symmetric
tensor, or $Sp$ with antisymmetric tensor:
\eqn\bj{\eqalign{
  SO/Sp(N)\to \prod_i G_i(N_i),
  \qquad
  G_i=U, SO, {\rm ~or~} Sp.
}}

The 3-form RR fluxes from orientifold hyperplanes makes an additional
contribution to the superpotential \bg, and the flux superpotential can be
written as
\eqn\bk{\eqalign{
  W_{\rm flux}
  =
  \sum_i [\Nh_i \Pi _i (S_i)- 2 \pi i \eta_i \alpha
S_i],
}}
where
\eqn\bl{\eqalign{
  \Nh_i=\cases{
    N_i         & $G_i=U(N_i)$, \cr
    \half N_i \mp 1 & $G_i=SO(N_i)/Sp(\half N_i)$, \cr
  }
  \qquad
  \eta_i =\cases{
    1   & $G_i=U$, \cr
    1/2 & $G_i=SO/Sp$. \cr
  }
}}
$\Nh_i$ is the net 3-form RR flux through the $A_i$ cycle.  For $U(N_i)$
and $Sp(N_i)$, $\Nh_i$ in \bl\ is the dual Coxeter number \wlowuiv,
while for $SO(N_i)$ it is half\foot{So we get $h$ replaced with $h/2$
for $SO$ groups in \wgcisi.  While one could absorb the overall factor
of 2 into the definition of $\Lambda$, the number of vacua should be $h$
whereas here we apparently get $h/2$ for $SO$ groups.  This is because
the we don't see spinors or the ${\bf Z}_2$ part of the center which
acts on them; it's analogous to $U(2N)$ being restricted to vacua with
confinement index 2.}  the dual Coxeter number \wlowuiv.  The $1/2$ in
\bl\ is because the integration over the $A_i$ cycles should be halved
due to the $\Ztwo$ identification.

\subsec{Relations between $SO/Sp$ theories and $U(N)$ theories}

The result \bk, with \bl, gives the exact superpotential of the $SO/Sp$
theories in terms of the same periods $S_i$ and $\Pi (S_i)$ as an
auxiliary $U$ theory.  This was first noted in \CachazoKX\ at the level
of the Konishi anomaly equation as a map between the resolvents of $Sp$
theory with antisymmetric matter and $U$ theory with adjoint matter. In
\LandsteinerPH, it was generalized to the map between the resolvents of
$SO/Sp$ theories with two-index tensor matter and $U$ theory with
adjoint matter, and string theory interpretation was discussed.
In this subsection we will derive this map from the string theory
perspective using the flux superpotential \bk.  Furthermore, we will
clarify the relation of the superpotential and the scale of the $SO/Sp$
theories to those of the $U$ theory.  The map between resolvents can be
derived from these results.  For $Sp(N)$ theory with an antisymmetric
tensor, the scale relation was obtained in a different way in
\CachazoKX.

As a first example, consider $SO(N)$ with an
adjoint, with the breaking pattern as in \higgscases.  The geometric
transition result \bk\ and \bl\ implies that the exact superpotential is
the same as for the $U(N-2)$ theory with adjoint, with breaking pattern
map
\eqn\soaaux{\eqalign{
  SO(N)\rightarrow SO(N_0)\times \prod _{i=1}^KU(N_i)\quad
  \iff \quad
  U(N-2)\rightarrow U(N_0-2)\times \prod_{i=1}^K U(N_i)^2.
}}
The map between the superpotential is
\eqn\soaauxsp{\eqalign{
  W^{SO(N)}_{\rm exact}=\half W^{U(N-2)}_{\rm exact}.
}}
%
The
$SO(N)$ scale matching relation \lamsoa\ is compatible with the map 
\soaaux, since \lamin\ for the theory \soaaux\ reproduces \lamsoa.

Likewise, $Sp(N)$ with adjoint has the same exact superpotential
as for the $U(2N+2)$ theory with adjoint, with
\eqn\spaaux{\eqalign{
  Sp(N)\to Sp(N_0)\times \prod _{i=1}^K U(N_i)\quad
  &\iff \quad
  U(2N+2)\rightarrow U(2N_0+2)\times \prod _{i=1}^K U(N_i)^2,\cr
  W^{Sp(N)}_{\rm exact}&=\half W^{U(2N+2)}_{\rm exact}.
}}
The $Sp(N)$ scale matching relations \lamspa\ follow from the $U(N)$
matching relations \lamin\ with the replacement \spaaux, with the
understanding that the $U(2N+2)$ and $U(2N_0+2)$ one-instanton factors
correspond to the $Sp(N)$ and $Sp(N_0)$ two-instanton factors; this is
related to the index of the imbedding mentioned after \lamspa, and is
accounted for by dividing the $U(2N+2)$ superpotential by two, as above.

Next consider $Sp(N)$ with antisymmetric tensor $A$ and breaking pattern
as in \higgscases.  The geometric transition result \bk\ and \bl\
implies that the exact superpotential is the same as for the $U(2N+2K)$
theory with adjoint and breaking pattern
\eqn\spasaux{\eqalign{
Sp(N)\rightarrow \prod _{i=1}^K Sp(N_i)\quad \iff \quad
U(2N+2K)\rightarrow \prod _{i=1}^K U(2N_i+2).
}}
In the present case, comparing the matching relations \laminsp\ for
$Sp(N)$ with symmetric tensor with the matching relations \lamin\ for
$U(2N+2K)$ with adjoint, for the mapping as in \spasaux\ requires that
the scales of the original $Sp(N)$ on the LHS of \spasaux\ and the
$U(2N+2K)$ on the RHS of \spasaux\ be related as
\eqn\spausr{\Lambda _{U(2N+2K)}^{2(N+K)}=g_{K+1}^{-2}\Lambda _{Sp(N)}^{2N+4}.}
Then the $\Lambda _i$ of the unbroken groups on both sides of \spasaux\
coincide, with the understanding that the $U(2N_i+2)$ one-instanton
factors correspond to the $Sp(N_i)$ two-instanton factors, as above.
The map between the superpotential is
\eqn\spausrsp{\eqalign{
  W^{Sp(N)}_{\rm exact}=\half [W^{U(2N+2K)}_{\rm exact}-\Delta W_{\rm cl}],
  \qquad&
  \Delta W_{\rm cl}=2\sum_{i=1}^K W(a_i),
  \cr
\hbox{i.e.\ writing}\quad W_{\rm exact}=W_{\rm cl}+W_{\rm quant},
 \qquad  W_{\rm quant}^{Sp(N)}=&
\half W_{\rm quant}^{U(2N+2K)}.
}}
Note that, in order for the superpotentials on the two sides of
\spasaux\ to fully coincide, one must compensate the classical mismatch
$\Delta W_{\rm cl}$, since each well is occupied by two additional
eigenvalues in the theory on the RHS of \spasaux.  In the string theory
geometric transition realization, this constant shift, which is
independent of $N$, $\Lambda$, and the glueball fields $S_i$, is most
naturally interpreted as an additive shift of the superpotential on the
$Sp$ side, which can be regarded as coming from the orientifold planes
both before and after the transitions.  The classical shift of $\Delta
W_{\rm cl}$ leads to slightly different operator expectation values (as
computed via $W_{\rm eff}(g_p, \Lambda)$ as the generating function)
between the $Sp$ and $U$ theory, as was seen in the example of
\CachazoKX.  Also, writing the map as in \spasaux, we want the vacuum
with confinement index 2 \CachazoKX.  We could equivalently replace the
RHS of \spasaux\ with $U(N+K)\rightarrow \prod _{i=1}^KU(N_i+1)$, in
which case we would not have to divide by 2 in \spausrsp.

Likewise, $SO(N)$ with symmetric tensor $S$ has exact superpotential related to 
that of a $U(N-2K)$ theory with adjoint as
\eqn\sosaux{\eqalign{
SO(N)\rightarrow \prod _{i=1}^K SO(N_i)\quad
  \iff& \quad U(N-2K)\rightarrow \prod _{i=1}^K U(N_i-2),\cr
  W^{SO(N)}_{\rm exact}=\half [W^{U(N-2K)}_{\rm exact }+\Delta W_{\rm cl}],
  &\qquad
  \Delta W_{\rm cl}=2\sum_{i=1}^K W(a_i).
  \cr
}}
Comparing the matching relations \lamsos\ for $SO(N)$ with symmetric
tensor with those of \lamin\ for $U(N-2K)$ with adjoint requires that
the scales of the original $SO(N)$ on the LHS of \sosaux\ and those of
the $U(N-2K)$ theory on the RHS be related as
\eqn\sosusr{
\Lambda_{U(N-2K)}^{2(N-2K)}=g_{K+1}^4\Lambda _{SO(N)}^{2N-8}.
}
Then the
$\Lambda _i$ of the unbroken groups on both sides of \sosaux\ coincide.
Again, in order for the superpotentials on the two sides of \sosaux\ to
fully coincide, one must correct for the classical mismatch $\Delta
W_{\rm cl}$ coming from the fact that the $U(N-2K)$ theory has two fewer
eigenvalues in each well.

In appendix A we will discuss these relations from the matrix model
viewpoint.  In this context, the relation relevant for \soaaux\ and
\spaaux\ was conjectured in \refs{\ItaKX,\AshokBI} based on explicit
diagrammatic calculations, and it was proven for the case of unbroken
gauge group $N_0=N$ in \JanikNZ.  This will be generalized in Appendix A
to all breaking patterns.  Likewise, the matrix model relation relevant
for \spasaux\ and \sosaux\ will be proven in the appendix; this is a
generalization of the connection found in \CachazoKX\ for the theories
in \spasaux\ and \sosaux.

\newsec{String theory prescription for low rank}

The discussion of the previous section applies for all $N_i\geq 0$.  We
now discuss under which circumstances one expects a transition in string
theory, where $S_i^3$'s grows, and therefore an effective glueball field
$S_i$ should be included in the superpotential.  Whether or not there is
a geometric transition in string theory is a local question, so each
$S_i^3$ can be studied independently.  Near any $S_i^3$ the local
physics is just a conifold singularity, so we only need to consider the
case of a conifold singularity.

\subsec{Physics near a conifold singularity}

As we saw, $U(N)$, $SO/Sp(N)$ gauge theory can be realized in type IIB
theory as the open string theory living on the D5-branes partially
wrapped on the exceptional $\Pone$ of a resolved conifold geometry.
There is a $\Pone$ associated to each critical point of the polynomial
$W(x)$.  By the geometric transition duality \refs{\VafaWI ,\CachazoJY},
this gauge theory is dual to the closed string theory in the deformed
conifold geometry where the $\Pone$'s have been blown down and $S^3$'s
are blown up instead.

Let us focus on one $\Pone$ with $N\ge 0$ D5-branes wrapping it.  This
corresponds to focusing on one critical point on the gauge theory side.
We allow $N=0$, which corresponds to an unoccupied critical point.  In
the neighborhood, the geometry after the geometric transition is
approximately a deformed conifold $x^2+y^2+z^2+w^2=\mu$ with a blown up
$S^3$.  The low energy degrees of freedom in the four-dimensional theory
are the $\CN=1$ $U(1)$ photon vector superfield $V$ and the $\CN=1$
chiral superfield $S$.  The bosonic component of $S$ is proportional to
$\mu$ and measures the size of the $S^3$.

%

First, consider the case {\it without\/}
fluxes.  Then the closed string theory has $\CN=2$ and there is one
$\CN=2$ $U(1)$ vector multiplet $(V,S)$.
It is known that as the size of the $S^3$ goes to zero there appears an
extra massless degree of freedom \StromingerCZ, which corresponds to the
D3-brane wrapping the $S^3$.  The mass of the wrapped BPS D3-brane is
proportional to the area, $S$, of the $S^3$, so the mass becomes zero as
the $S^3$ shrinks to zero, i.e.\ as $S\to0$.
This extra degree of freedom is described as an $\CN=2$ hypermultiplet
charged under the $U(1)$ (of $V$).  Let us write this hypermultiplet in
$\CN=1$ language as $(Q,\Qt)$, where $Q$ and $\Qt$ are both $\CN=1$
chiral superfields with opposite $U(1)$ charges.  The $\CN=2$
supersymmetry requires the superpotential
\eqn\ca{\eqalign{
W_Q = \sqrt{2}\,Q\Qt S,
}}
which indeed incorporates the above situation that the $Q,\Qt$ become
massless as $S\to 0$.  The $D$-flatness is
\eqn\cb{\eqalign{
|Q|^2 - |\Qt|^2 = 0 ,
}}
and the $F$-flatness is
\eqn\cc{\eqalign{
QS=\Qt S=Q\Qt=0.
}}
The only solution to these is
\eqn\cd{\eqalign{
Q=\Qt=0, \qquad S:{\rm any},
}}
which just means that $S\sim \mu$ is a modulus.
%

Now let us come back to the case with the fluxes.  As reviewed in the
last section, the fluxes give rise to a superpotential \bg\ which breaks
$\CN=2$ to $\CN=1$ \refs{\TaylorII ,\MayrHH}.  As in \bk, the local flux
superpotential contribution is 
\eqn\ce{\eqalign{
  W_{\rm flux}(S)\simeq \Nh S[1-\ln(S/\Lambda^3)] - 2\pi i\eta \alpha S,
}}
where we just keep the leading order term in \bk, as in \bi, with 
\eqn\cg{\eqalign{
  \Nh=\cases{
    N & $U(N)$, \cr
    N/2 \mp 1 & $SO(N)/Sp(\half N)$, \cr
  }
  \qquad
  \eta =\cases{
    1 & $U(N)$, \cr
    1/2 & $SO(N)/Sp(\half N)$. \cr
  }
}}
The scale $\Lambda$ is written in terms of the bare coupling $\Lambda_b$
and the coupling constants in the problem, as before, and $2\pi i
\alpha$ is related to the bare gauge coupling by \bha.

In the following, we discuss the cases with $\Nh=0$, $\Nh>0$ and $\Nh<0$
in order.
%
\medskip
\item{$\bullet$} {\bf \^N=0 case}\hfil\break
In this case, the total superpotential is simply the sum of \ca\ and
\ce:
\eqn\ci{\eqalign{
  W=\sqrt{2}\,Q\Qt S-2\pi i\eta \alpha S.
}}
The only solution to the equation of motion is
\eqn\ck{\eqalign{
   |Q|^2=|\Qt|^2,\qquad Q\Qt={2\pi i\eta \alpha\over \sqrt{2}},\qquad S=0.
}}
This is consistent with the fact that $\alpha$ is proportional to the
volume of the $\Pone$, and the D3-brane condensation $\langle Q\Qt
\rangle $ corresponds to the size of the $\Pone$.  Furthermore, since
$\ev{S}=0$, the superpotential vanishes: $W=0$.  Therefore, for $\Nh=0$,
i.e.\ for $U(0)$ and $SO(2)$, geometric transition does not take place
and we should set the corresponding glueball field $S\rightarrow 0$ from
the beginning.

%
\medskip
\item{$\bullet$} {\bf \^N$>$0 case}\hfil\break
In this case, there is a net RR flux through the $A$-cycle: $\oint_A
H=\Nh$.  This means that the D3-brane hypermultiplet $(Q,\Qt)$ is
infinitely massive, because the RR flux will induce $\Nh$ units of
fundamental charge on the D3-brane.  Since the D3-brane is wrapping a
compact space $S^3$, the fundamental charge on it should be canceled by
$\Nh$ fundamental strings attached to it.  Those fundamental strings
extend to infinity and thus cost infinite energy\foot{This phenomenon is
the same as that observed in the context of AdS/CFT \refs{\WittenXY ,
\GrossGK , \GubserFP}.}. Therefore, we can forget about $Q,\Qt$ in this
case, and the full superpotential is given just by the flux contribution
\ce.  The equation of motion gives
\eqn\cl{\eqalign{
   S^{\Nh} \simeq \Lambda^{3\Nh} e^{-2\pi i\eta \alpha},
}}
which corresponds to the confining vacua of the gauge theory. Note that
this case includes $U(1)$ and $Sp(0)$; these theories have a dual
confining description.  This may sound a little paradoxical, but is
related to the fact that the string theory computes not for the standard
$G(N)$ gauge theory but the associated $G(N+k|k)$ higher rank gauge
theory, which is confining and differs from standard $U(1)$ and $Sp(0)$
due to residual instanton effects \AIVW.

%
\medskip
\item{$\bullet$} {\bf \^N$<$0 case}\hfil\break
In this case, the same argument as the $\Nh>0$ case tells us that we
should not include the D3-brane fields $Q,\Qt$.  Hence the
superpotential is just the flux part \ce, which again leads to
\eqn\cm{\eqalign{
   S^{\Nh} \simeq \Lambda^{3\Nh} e^{-2\pi i\eta \alpha}.
}}
However, now \cm\ is physically unacceptable, since $S$ diverges in the
weak coupling limit where the bare volume of $\Pone$ becomes $V=2\pi i
\alpha\to \infty$ ($g_b\to 0$) -- i.e.\ taking $\Pone$ large would lead
to $S^3$ also being large, which does not make sense geometrically.  The
resolution is that $S$ cannot be a good variable: the $S^3$ does not
actually blow up, and $S$ should be set to zero, $S \rightarrow 0$, also
for this case.  Though $S$ is set to zero, the non-zero flux can lead to
a non-zero superpotential contribution $W_{\rm flux}=\Nh \Pi
(S\rightarrow 0)$.

\bigskip
Note that the above result concerning the sign of $\Nh$ does
not mean the gauge theory prefers D5-branes to anti-D5-branes; it just
means that one should choose the sign of the NS flux (i.e.\ the sign of
$2\pi i \alpha$) appropriately.  If one wraps the $\Pone$ with
anti-D5-branes, one should flip the sign of the NS flux in order to have
a blown up $S^3$ (which can be viewed as a generalization of Seiberg
duality to $N_f=0$).

\subsec{General prescription}

Although we focused on the physics around just one $\Pone$ in the above,
the result is applicable to general cases where we have multiple $\Pone$
wrapped with D5-branes, because the geometry near each $\Pone$ is
identical to the conifold geometry considered above.  Therefore, if we
replace $\Nh$ with $\Nh_i$, all of the above conclusions carry over.

Once we have understood the physics, we can forget about the D3-brane
hypermultiplet $(Q,\Qt)$ and state the result as a general prescription
for how string theory treats $U(0)$, $SO(0)$, $SO(2)$, and $Sp(0),U(1)$
groups in the geometric dual description: \medskip
\item{$\bullet$}
{\bf U(0), SO(0), SO(2):}
\hfil\break
There are no glueball variables associated
to these gauge groups, so we should take the corresponding $S\rightarrow 0$.
\medskip
\item{$\bullet$}
{\bf All other groups, including Sp(0), and U(1):}
\hfil\break
We should consider and extremize the corresponding glueball field $S$.
\medskip

This prescription should also be applied when using the matrix
model \refs{\DijkgraafFC ,\DijkgraafVW  ,\DijkgraafDH}\ to compute the glueball
superpotentials.

\newsec{Examples}

Let us scan over all of the examples of \higgscases, considering the
vacuum where the gauge group is unbroken, and ask when glueball fields
$S_i$ for the apparently trivial groups in \higgscases\ should be set to
zero, or included and extremized.  For the first three cases in
\higgscases, $U(N)$, $SO(N)$, and $Sp(N)$ with adjoint, the breaking
\higgscases\ is $G\rightarrow G\times U(0)^{K-1}$, and the glueball
fields $S_i$ for the $U(0)$ factors are to be set to zero.  This
justifies the analysis of these theories in the unbroken vacua in
\refs{\FerrariJP,\JanikNZ}.
The next case is $SO(N)$ with a symmetric tensor $S$, where the vacuum
with unbroken gauge group is to be understood as $SO(N)\rightarrow
SO(N)\times SO(0)^{K-1}$, and again the glueball fields $S_i$ for the
$SO(0)$ factors are set to zero.  This eliminates the
Veneziano-Yankielowicz part of the superpotential for $SO(0)$, but the
$-1$ unit of flux associated with each $SO(0)$ does contribute to flux
terms $\Nh_i \Pi _i=-\Pi_i$ in \bk, even though this does not contain
$SO(0)$ glueballs any more.

The next case is $Sp(N)$ with an antisymmetric tensor, where the vacuum
with unbroken gauge group is to be understood as $Sp(N)\rightarrow
Sp(N)\times Sp(0)^{K-1}$.  Unlike the above cases, here we must keep and
extremize the $S_i$ for the $Sp(0)$ factors, as will be further
discussed shortly.

For the next to last example in \higgscases, $U(N)$ with $\Phi
+S+\widetilde S$, the vacuum with unbroken gauge group is to be
understood as $U(N)\rightarrow SO(0)\times U(N)\times U(0)^{K-1}$, and
the glueball fields $S_i$ for $SO(0)$ and $U(0)$ are to be set to zero.
Finally, for the last example in \higgscases, $U(N)$ with $\Phi
+A+\widetilde A$, the vacuum with unbroken gauge group is to be
understood as $U(N)\rightarrow Sp(0)\times U(N)\times U(0)^{K-1}$.
Though the string engineering of these examples differs somewhat from
those discussed in sect.\ 4 (it was obtained in \LandsteinerRH), the
general prescription of sect.\ 4 is expected to carry over in general:
the glueball field $S_0$ for the $Sp(0)$ factor should be included and
extremized, rather than set to zero.  On the other hand, the $S_i$ for
the $U(0)$ factors are set to zero.  These latter two theories in
\higgscases\ were considered in \NaculichKA\ and it was noted there that
for the case with antisymmetric one expands on the matrix model side
around a different vacuum than would be naively expected; this indeed
corresponds to keeping and extremizing the glueball field $S_0$ for the
$Sp(0)$ factor, as we have discussed.

We now illustrate some other breaking patterns in the examples of
\higgscases, from the matrix model perspective, for the case of $K=2$.
We also compare with standard gauge theory results and generally find
agreement, even in cases where there was room for disagreement because
of the possibility of residual instanton effects along the lines of
\AIVW.  As will be discussed in more detail in the following section,
the agreement is thanks to a remarkable interplay of different residual
instanton contributions.

\subsec{$SO/Sp(N)$ theory with adjoint} Consider $\CN=2$ $SO(2N)/Sp(N)$
theory broken to $\CN=1$ by a tree level superpotential for the adjoint
chiral superfield $\Phi$:
\eqn\da{\eqalign{
  W_{\rm tree}=\half \Tr[W(\Phi)],\qquad
  W(x)={m\over 2}x^2+{g\over 4}x^4.
}}
In the $SO$ case, we can skew-diagonalize $\Phi$ as
\eqn\db{\eqalign{
\Phi
&\sim{\rm diag}[\lambda_1,\cdots,\lambda_{N}]\otimes i\sigma^2.
}}
The superpotential \da\ has critical points at $\lambda=0$ and
$\lambda=\pm\sqrt{m/ g}$.  The classical supersymmetric vacuum of the
theory is given by distributing $2N_0$ of the $2N$ ``eigenvalues''
$\lambda_i$ at the critical point $\lambda=0$ and $N_1$ ``eigenvalue''
pairs at $\lambda=\pm \sqrt{m/g}$, with $N_0+N_1=N$.  In this vacuum,
the gauge group breaks as $SO(2N)\to SO(2N_0)\times U(N_1)$ or $Sp(N)\to
Sp(N_0)\times U(N_1)$.

In the matrix model prescription, the effective superpotential in these
vacua is calculated by matrix model as
\eqn\dc{\eqalign{
  W_{\rm DV}(S_0,N_0;S_1,N_1)
  &=
  \left(N_0 \mp 1\right)S_0[1-\ln(S_0/\Lambda_0^3)]
  +N_1S_1[1-\ln(S_1/\Lambda_1^3)]+W_{\rm pert},
  \cr
  W_{\rm pert}
  &=
  2N_0{\partial \CF_{S^2}\over \partial S_0}
  + N_1{\partial \CF_{S^2}\over \partial S_1} + 4\CF_{RP^2}.
}}
where $\CF_{S^2}$ and $\CF_{RP^2}$ are the $S^2$ and $RP^2$
contributions, respectively, to the free energy of the associated
$SO/Sp(N)$ matrix model, as defined in Appendix A.  The scales
$\Lambda_0$, $\Lambda_1$ in \dc\ are the energy scales of the low energy
$SO(2N_0)/Sp(N_0)$, $U(N_1)$ theories with the $\Phi$ field integrated
out, respectively.  They are related to the high energy scale $\Lambda$
by the matching conditions as in \lamsoa\ and \lamspa, which yields
\eqn\de{\eqalign{
  (\Lambda_0)^{3(N_0\mp 1)}
  &=
  m^{N_0-N_1\mp 1} g^{N_1}
  \Lambda^{2(N \mp 1)}
  ,\cr
  (\Lambda_1)^{3N_1}
  &=2^{-N_1}
  m^{-2N_0\pm 2} g^{2N_0+N_1\mp 2} \Lambda^{4(N \mp 1)} .
}}

The matrix model free energy is computed in Appendix A, and the result
is
\eqn\dd{\eqalign{
  W_{\rm pert}
  =&
  \left({N_0} \mp 1\right)
  \biggl[
    \left( {3\over 2}S_0^2-8S_0S_1+2S_1^2\right) \alpha
    + \left( -{9\over 2}S_0^3 + 42S_0^2S_1 - 36S_0S_1^2 + 4S_1^3\right)
\alpha^2
    \cr
    &
    \qquad\qquad\qquad
    +\left(
      {45\over 2}S_0^4 - {932\over 3}S_0^3S_1 + 523S_0^2S_1^2
      -{608\over 3}S_0S_1^3 + {40\over 3}S_1^4
    \right) \alpha^3
  \biggr]
  \cr
  &+{N_1}
  \biggl[
    \left( -2S_0^2+2S_0S_1 \right) \alpha
    + \left( 7S_0^3 -18S_0^2S_1+6S_0S_1^2 \right) \alpha^2
  \cr
  &
  \qquad\qquad\qquad
  + \left(
      -{233\over 6}S_0^4 + {524\over 3}S_0^3S_1
      - 152S_0^2S_1^2 + {80\over 3}S_0S_1^3
    \right) \alpha^3
  \biggr]
  +\CO(\alpha^4),
}}
where $\alpha\equiv g/m^2$.  The result \dd\ agrees with the one
obtained in \FujiVV, where the glueball superpotential was calculated by
evaluating the periods \bg.  The full result, \dc\ and \de, has the
expected general form \bk:
\eqn\ddd{W_{\rm eff}=(N_0\mp 1) \Pi _0 (S_0, S_1)+N_1\Pi _1(S_0, S_1)-2\pi i 
\alpha (\half S_0+S_1).}

The general prescription in section 3 reads in the present
case as follows:
\medskip
\item{$\bullet$}
{\bf SO(2N)/Sp(N)$\,\to\,$SO(2N)/Sp(N)$\times$U(0) (unbroken SO/Sp):}
\hfil\break
Set $N_1=0$, $S_1=0$.  Then the superpotential is
\eqn\df{\eqalign{
  W_{\rm eff}(S_0,N_0)
  &=
  \left({N_0} \mp 1\right)S_0[1-\ln(S_0/\Lambda_0^3)]
  +2\left(N_0 \mp 1\right)
  \left.{\partial \CF_{S^2}\over \partial S_0}\right|_{S_1=0}.
}}
This superpotential coincides with that of $U(2N\mp 2)$ with adjoint and
breaking pattern $U(2N\pm 2)\rightarrow U(2N\mp 2)\times U(0)\times
U(0)$, as expected from the map \soaaux\ or \spaaux.  As shown in
\JanikNZ\ for $SO(2N)$, this matrix model result agrees with that of
standard gauge theory, via using the corresponding Seiberg-Witten curve.
\medskip
\item{$\bullet$}
{\bf SO(2N)$\,\to\,$SO(0)$\times$U(N):}
\hfil\break
Set $N_0=0$ and take $S_0\rightarrow 0$, which eliminates the
Veneziano--Yankielowicz part for the $SO(0)$.  Then the superpotential
is
\eqn\dg{\eqalign{
  W_{\rm eff}(S_1)
  &=
  NS_1[1-\ln(S_1/\Lambda_1^3)]
  +4\left.\CF_{RP^2}\right|_{S_0=0}.\cr
&=-\Pi _0(S_0, S_1)|_{S_0=0}+N\Pi _1(S_0, S_1)|_{S_0=0}-2\pi i \alpha S_1.
}}
Note that $\left.{\partial \CF_{S^2}\over \partial
S_1}\right|_{S_0=0}=0$ because $\CF_{S^2}$ does not contain terms with
$S_1$ only (all terms are of the form $S_0^nS_1^m$ with $n>0$).  And
though the Veneziano-Yankielowicz part of the superpotential for $SO(0)$
is eliminated via $S_0\rightarrow 0$, the $-1$ units of flux associated
with $SO(0)$ does make a contribution in \dg, with the non-zero terms in
$-\Pi _0(S_0, S_1)|_{S_0=0}$ in the second line of \dg\ coming from the
term $4\CF _{RP^2}|_{S_0=0}$.

\medskip
\item{$\bullet$}
{\bf SO(2N)$\,\to\,$SO(2)$\times$U(N$-$1):}
\hfil\break
Set $N_0=1$, $S_0=0$ and remove the Veneziano--Yankielowicz part for the
$SO(2)$.  Then the superpotential is
\eqn\dh{\eqalign{
  W_{\rm DV}(S_1) &= (N-1)S_1[1-\ln(S_1/\Lambda_1^3)],
}}
where we used $\left.{\partial \CF_{S^2}\over \partial
S_1}\right|_{S_0=0}=0$ again.  Integrating out $S_1$ gives
\eqn\di{\eqalign{
  W_{\rm low}
  =(N-1)\Lambda_1^3 =\half (N-1)g\Lambda^4 }}
\item{$\bullet$}
{\bf Sp(N)$\,\to\,$Sp(0)$\times$U(N):}
\hfil\break Set $N_0=0$ in the equation and keep both $S_0$ and $S_1$.
Then the superpotential is
\eqn\dk{\eqalign{
  W_{\rm DV}(S_0,N_0,S_1,N_1)
  &=
  S_0[1-\ln(S_0/\Lambda_0^3)] + NS_1[1-\ln(S_1/\Lambda_1^3)]+W_{\rm
pert},\cr
  W_{\rm pert}
  &=
  N_1{\partial \CF_{S^2}\over \partial S_1} + 4\CF_{RP^2}
  =
  N_1{\partial \CF_{S^2}\over \partial S_1}
  +2{\partial \CF_{S^2}\over \partial S_0}.
}}
%
%
%
{
\par\begingroup\parindent=0pt\leftskip=1cm\rightskip=1cm\parindent=0pt
\baselineskip=11pt
\global\advance\figno by 1
\midinsert
\epsfxsize=10cm
\centerline{\epsfbox{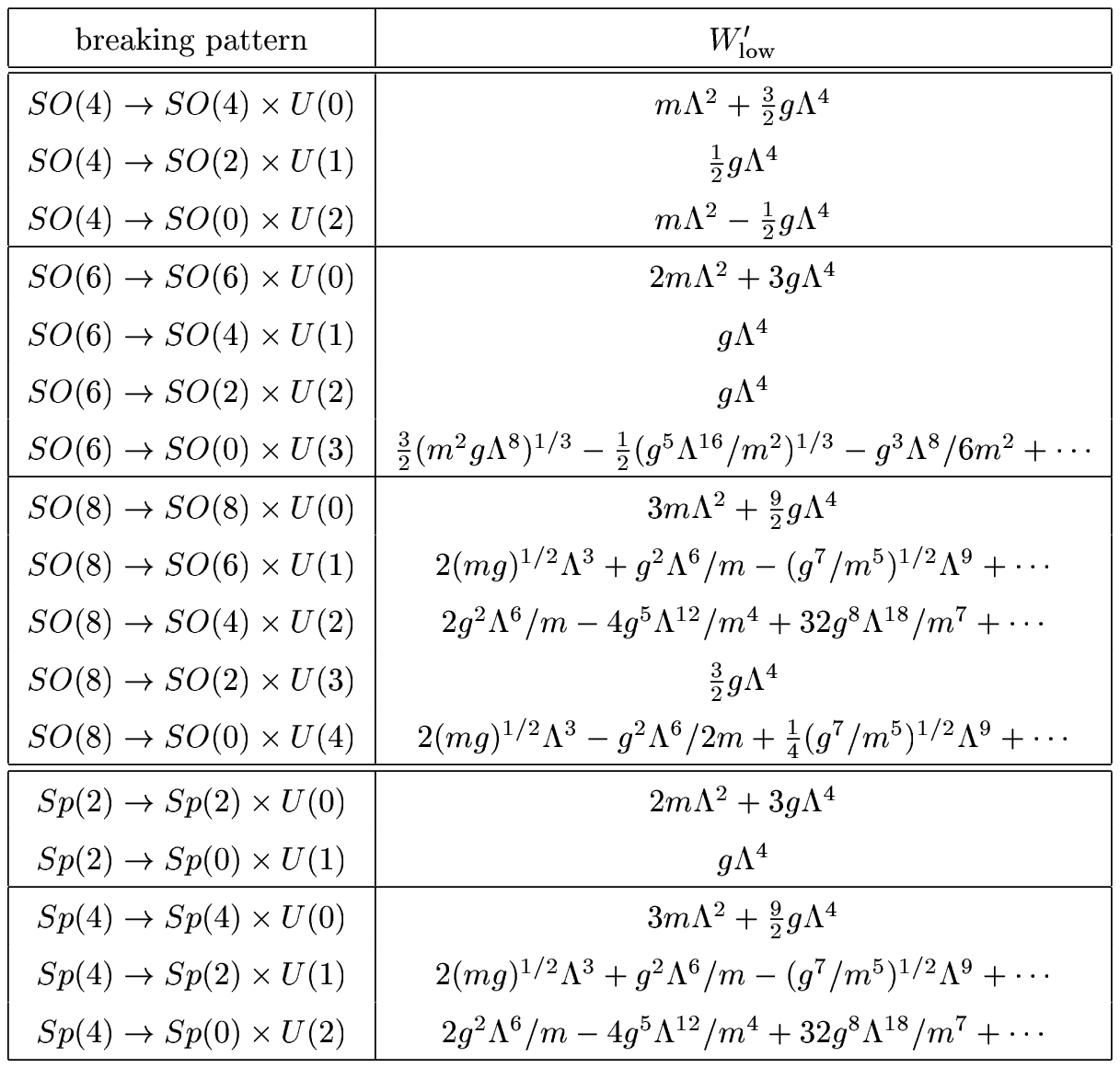}}
\vskip 10pt
\footnotefont
{\bf Table \the\figno: }
The low energy superpotential calculated from the factorization of the
Seiberg-Witten curve and from matrix model.  In the above, the 
classical contribution has been subtracted:
$W_{\rm low}=-{N_1 m^2/ 4g}+W'_{\rm low}$.
\par
\endinsert\endgroup\par
}
\figlabel{\superpottable}

For various breaking patterns, we integrated out the glueball
superfield(s) from the glueball superpotential \df--\dk, and calculated
the low energy superpotential $W_{\rm low}$ as a function of coupling
constants $m$, $g$, and the scale $\Lambda$.  Having obtained the actual
matrix model results, we can compare to the superpotential as computed
via standard gauge theory methods, such as via factorizing of the
Seiberg--Witten curve.  This method is reviewed in Appendix B, and the
results are found to agree with the matrix model results completely.
The resulting $W_{\rm low}$ is shown in Table \superpottable.

The $SO(2N+1)$ theory with adjoint in the $SO(2N+1)\to U(1)^N$ vacuum
was studied diagrammatically in \AbbaspurHF.

\subsec{$Sp(N)$ theory with antisymmetric tensor}

Consider $Sp(N)$ theory with an antisymmetric tensor chiral superfield
$A=-A^T$.  Take cubic tree level superpotential
\eqn\ea{\eqalign{
  W_{\rm tree}=\half \Tr[W(\Phi)],\qquad
  W(x)={m\over 2}x^2+{g\over 3}x^3,
}}
where $\Phi=AJ$, and $J$ is the invariant antisymmetric tensor $J={\bf
1}_{N} \otimes i\sigma^2$.  We do not require $A$ to be traceless, i.e.\
$\Tr[\Phi]=\Tr[AJ]\neq 0$.  By a complexified $Sp(N)$ gauge rotation,
$\Phi$ can be diagonalized as \ChoBI
\eqn\ec{\eqalign{
  \Phi
  &\cong  {\rm diag}[\lambda_1,\cdots,\lambda_{N}]\otimes {\bf 1}_2,
\qquad
  \lambda_i\in {\bf C}.
}}
The superpotential \ea\ has critical points at $\lambda=0,-m/g$.  The
classical supersymmetric vacuum of the theory is given by distributing
$N_1$ and $N_2$ ``eigenvalues'' $\lambda_i$ at the critical point
$\lambda=0$ and $\lambda=-{m/g}$, respectively, with $N_1+N_2=N$,
breaking $Sp(N)\to
Sp(N_1)\times Sp(N_2)$.

The glueball superpotential is calculated from the associated $Sp(N)$
matrix model as
\eqn\ed{\eqalign{
  W_{\rm DV}(S_1,N_1;S_2,N_2)
  =&
  ({N_1} + 1)S_1[1-\ln(S_1/\Lambda_1^3)]\cr
  &\qquad\qquad
  +({N_2} + 1)S_2[1-\ln(S_2/\Lambda_2^3)]
  +W_{\rm pert},
  \cr
  W_{\rm pert}
  &=
 2 N_1{\partial \CF_{S^2}\over \partial S_1}
  + 2N_2{\partial \CF_{S^2}\over \partial S_2} + 4\CF_{RP^2}.
}}
The scales
$\Lambda_1$, $\Lambda_2$ in \dc\ respectively are the energy scales of
the low energy $Sp(N_1)$, $Sp(N_2)$ theories with the $\Phi$ field
integrated out.  They are related to the high energy scale $\Lambda$ by
the matching conditions as in \laminsp, which yields
\eqn\ef{\eqalign{
  (\Lambda_1)^{3(N_1+1)}
  &= m^{{N_1}-2N_2-1} g^{2N_2}
\Lambda^{2N+4} ,\cr
  (\Lambda_2)^{{3}(N_2+1)}
  &=
  (-1)^{{N_2}-1}
   m^{-2N_1+{N_2}-1} g^{2N_1}
\Lambda^{2N+4}.
}}

The matrix model free energy is computed in Appendix A, and the result
is
\eqn\eg{\eqalign{
 W_{\rm pert}
 =&\,
 2(N_1+1)
 \biggl[
   \left( -{S_1^2} + 5 S_1 S_2 - {5\over 2}S_2^2 \right)\alpha
   +
   \left(
    -{16\over 3}S_1^3 + {91\over 2}S_1^2 S_2 - {59}S_1 S_2^2
    + {91\over 6} S_2^3
   \right)\alpha^2
   \cr
   &\qquad\qquad+
   \left(
    -{140\over 3}S_1^4
    + {1742\over 3}S_1^3 S_2
    - {1318}S_1^2 S_2^2
    + {2636\over 3}S_1 S_2^3
    - {871\over 6} S_2^4
   \right)\alpha^3
 \biggr]
 \cr
 &+2(N_2+1)\biggl[S_1\leftrightarrow S_2,\, \alpha\to -\alpha\biggr],
 \cr
 }}
where $\alpha\equiv g^2/m^3$.  This is as expected from the map \spasaux\ and
\CachazoKX.

In particular, let us concentrate on the unbroken case, $N_2=0$.  Unlike 
\refs{\KrausJF, \KrausJV}, we do not set $S_2=0$, but rather 
keep $S_2$ non-zero and extremize with respect to it, according to our general
prescription, to obtain the actual matrix model result.  
After integrating out $S_1$ and $S_2$ from \eg, we
obtain the superpotential as a power series in $\Lambda_1$ and
$\Lambda_2$ as
\eqn\eh{\eqalign{
  W_{\rm low}= ( N+1 ) \Lambda_1^3+\Lambda_2^3
  +({\rm higher~order~terms~in~}\Lambda_{1,2}).
}}
The matching relation \ef\ gives
\eqn\ei{\eqalign{
  \Lambda_2^3=-(\Lambda_1^3)^{{N}+1}\alpha^{{N}},
}}
so the terms containing $\Lambda_2$ in \eh\ starts to contribute
to the superpotential at order $(\Lambda_1^3)^{{N}+1}$, i.e.\ like $Sp(N)$
instantons. 
If we use the relation \ei\ and write out all the
terms in \eh, we obtain
\eqn\ej{\eqalign{
  N=0: & \quad W_{\rm low}=\CO(\alpha^4), \cr
  N=1: & \quad W_{\rm low}=2\Lambda_1^3 + \CO(\alpha^4),\cr
  N=2: & \quad W_{\rm low}=3\Lambda_1^3 - \Lambda_1^6 \alpha
          -2 \Lambda_1^9 \alpha^2 -{187\over 27} \Lambda_1^{12}\alpha^3
    +\CO(\alpha^4), \cr
  N=3: & \quad W_{\rm low}=4\Lambda_1^3 - 3\Lambda_1^6 \alpha
          -{47\over 6} \Lambda_1^9 \alpha^2
    -{75\over 2} \Lambda_1^{12}\alpha^3
    +\CO(\alpha^4), \cr
  N=4: & \quad W_{\rm low}=5\Lambda_1^3 - 5\Lambda_1^6 \alpha
          -{13} \Lambda_1^9 \alpha^2
    -{65} \Lambda_1^{12}\alpha^3
    +\CO(\alpha^4). \cr
}}
Thus properly accounting for $S_2$, it turns out that these
matrix model results agree perfectly, up to the order presented,
with the standard gauge theory results (Eq.\ (4.13)
of \KrausJF).   (The
discrepancies  found in \KrausJF\ set in at order $\Lambda_1^{3(N+1)}$, 
and are cancelled e.g.\ by \ei.)
In \ej, for $N=0,1$, there were rather remarkable cancellations between
the instanton contributions from $Sp(N_1)$ and $Sp(N_2)$.  This will be
further discussed
and generalized in the next section.

The matrix model prediction for the superpotential of the $SO(2N)$ theory with
symmetric tensor can similarly be obtained by simply changing the $N_i+1$
in \eg\ to $N_i-1$.  It should be possible to compute the superpotential from gauge
theory using the duality for this theory \IntriligatorFF.  The result is expected to be
compatible with the map \sosaux\ to the superpotential computed for the $U(N-2K)$
theory with adjoint. 

\newsec{Residual Instantons:  String theory (matrix model) versus
gauge theory}

A remarkable aspect of the string theory (matrix model) computation
of the effective superpotential is that \prepotgen\ can be obtained
purely in terms of the dynamics of the low-energy $\prod _{i=1}^K G(N_i)$ theory on
the RHS of \higgsbp.  The only information needed about the high-energy $G(N)$
gauge theory is the perturbative contribution of the
$G(N)/\prod _i G_i(N_i)$ ghosts to the glueball superpotential
\prepotgen, as discussed in \dgkv, along with the matching relations connecting
the scales $\Lambda _i$ of the low-energy $G_i(N_i)$ factors to the scale $\Lambda$
of the high-energy $G(N)$ theory.   This is very different from the conventional
description of standard gauge theory, where there can be non-perturbative
contributions to $W_{\rm low}$ which are not readily seen in terms of the low-energy
theory on the RHS of \higgsbp.  An example of such an effect is instantons in the broken part of the
group when $\pi _3 (G(N)/\prod _i G_i(N_i))\neq 0$ (see e.g. \CsakiVV).  
Nevertheless, the string theory/ matrix model does properly reproduce such effects, via a low-energy
description.

A gauge theory interpretation for the string theory/ matrix model
results was given in \refs{\DijkgraafXK , \AIVW}: the string theory /
matrix model results actually refer to a particularly natural UV
completion of the original $G(N)$ theory, where it is embedded in the
supergroup $G(N+k|k)$ with $k$ large.  This latter theory has a Higgs
branch, where $k$ can be reduced successively, eventually Higgsing the
theory down to the original $G(N)$ theory.  More generally, the theory with
breaking pattern \higgsbp\
is replaced with
\eqn\higgsbpsg{G(N+k|k)\rightarrow \prod _{i=1}^KG_i(N_i+k_i|k_i),}
which has a Higgs branch flat direction connecting it to \higgsbp.  Consideration
of the particular matter content of the $G(N+k|k)$ theories along the Higgs branch, 
which often has extended supersymmetry, suggests that no dynamically generated
superpotential ever lifts this Higgs branch moduli space, i.e.\ that the
superpotentials of these particular theories are always 
independent of the location of the theory on this Higgs
branch \AIVW.   Moving along the Higgs branch has the effect of reducing $k$, and
this expected independence of the superpotential of the position on the Higgs branch
fits with the fact that the $G(N+k|k)$ matrix model results are $k$ independent, because
all $k$ dependence cancels in the supertraces.    

Because of the expected independence
of the superpotential on the Higgs branch, and because we Higgs back to the original 
$G(N)$ theory,  in most cases, this
``F-completion'' of the original $G(N)$ theory into the $G(N+k|k)$
theory is of no consequence.  There are, however, a few rare exceptions,
where the superpotential of the Higgsed $G(N+k|k)$ theory differs from
that of the standard $G(N)$ theory.  This difference comes from 
residual instantons in $G(N+k|k)/G(N)$, which
need not decouple even if $G(N+k|k)$ is Higgsed to $G(N)$ far in the UV.  
As verified in
\AIVW, these residual instanton contributions precisely account for the
few differences between the string theory (matrix model) results and
standard gauge theory, for example the glueball superpotentials, with
coefficient $h=1$, for $U(1)$ and $Sp(0)$, e.g.\ with an adjoint and
quadratic superpotential.   

In many cases, however, these residual instanton contributions sum up to
yield precisely the result expected from standard gauge theory, including
superpotential contributions which in standard gauge theory would not have
had a known low-energy description. In particular, residual instanton contributions
which could have lead to potential discrepancies with standard
gauge theory often completely cancel.  The cancellation occurs 
once one sums over the different terms $i$ in
\weffgen, upon using the precise matching relation between the low-energy
scales $\Lambda _i$, and the original high-energy scale $\Lambda$.  

As an example, consider $U(K)$ with adjoint matter and breaking pattern
$U(K)\rightarrow U(1)^K$. For
$K=1$ the string theory (matrix model) description
includes a residual
instanton effect, yielding $W_{\rm low}=\Lambda _L^3$ rather than the
standard gauge theory
answer $W_{\rm sgt}=0$ \AIVW.  But for all $K>1$ the string theory/ matrix model
result is  $W_{\rm low}=0$,
in agreement with the standard gauge theory expectation for
$U(K)\rightarrow U(1)^K$.  The
result $W_{\rm low}=0$ looks like a remarkable cancellation because the
glueball superpotential $W_{\rm eff}(S_1, \dots S_K)$ is quite non-trivial.
Nevertheless, upon
solving for the $\ev{S_i}$ and plugging back in, the exact result for
$W_{\rm low}=W_{\rm eff}(\ev{S_i})$
is zero, as was proven in \CachazoPR.

To illustrate this cancellation, consider the leading order gaugino
condensation contribution
to $W_{\rm eff}(S_i)$ in the string theory (matrix model) constructions,
where the unbroken $U(1)$
factors in $U(K)\rightarrow U(1)^K$ contribute as in \wgca, with
$h_i=1$, unlike in standard gauge theory:
\eqn\wgcaa{W_{\rm gc}(S_i)=\sum _{i=1}^kS_i \left(\log({\Lambda _i^3\over
S_i})+1\right),}
with $\Lambda _i^3=g_{K+1}\Lambda ^{2N}\prod _{j\neq i} (a_j-a_i)^{-1}$
by using
\lamin\ with all $N_i=1$.  Though this is a non-trivial superpotential,
it vanishes upon
integrating out the $S_i$:
\eqn\wgcasv{W_{\rm gc}(\ev{S_i})=\sum _{i=1}^K \Lambda _i ^3=
g_{K+1}\Lambda ^{2N}\sum _{i=1}^K \prod _{j\neq i}
(a_j-a_i)^{-1}=g_{k+1}^2\Lambda ^{2N}\oint {dx\over 2\pi i}{1\over
W'(x)}=0.} The contour in \wgcasv\ encloses all the zeros of $W'(x)$,
and we get zero for all $K>1$ by pulling
the contour off to infinity.
We see here why $K=1$ is different: we then get a residue at infinity,
leading to the low energy superpotential $W_{U(1)}=\Lambda _L^3$, as in
\wgcisi,  with $h_{U(1)}=1$ as in \wlowuiv.

To give another example of such a cancellation of residual instanton
effects, consider the
string theory (matrix model) result for $Sp(N)$ with an antisymmetric
tensor $A$, with $W_{\rm tree}$
having $K$ critical points, for the case $N=0$.   For the case of $K=1$,
the superpotential is
just a mass term for $A$ and the low-energy superpotential is the
$Sp(0)$ gaugino condensation
superpotential, with $h(Sp(0))=1$:  $W=\Lambda ^3$, unlike standard
gauge theory.
Again, this can be understood as a residual instanton effect in the
F-completion of
$Sp(N)$ to $Sp(N+k|k)$, which is present precisely for the case $N=0$
\AIVW.  For a higher order superpotential, $K>1$, we would write
the breaking pattern as
$Sp(0)\rightarrow Sp(0)^K$.  For all $K>1$, the residual instanton
effects all cancel, precisely as
in the $U(K)\rightarrow U(1)^K$ example discussed above; in fact, the
two theories have the
same effective superpotential $W_{\rm eff}$ (aside from the classical
difference), as discussed in \CachazoKX\ and sect.\ 3.3.
Thus, for example, \wgcasv\ can also be interpreted as the leading
gaugino condensation contributions from the $Sp(0)^K$ factors,
and where we now use the matching relation \laminsp\ to relate the 
$\Lambda _i^3$ to $g_{K+1}\Lambda^{2K}\prod _{j\neq i}(a_j-a_i)^{-1}$.
Again, there is complete cancellation in $W_{\rm eff}$ here, except for
the case $K=1$.

More generally, for $Sp(N)$ with antisymmetric, breaking as
$Sp(N)\rightarrow \prod _{i=1}^KSp(N_i)$, the results obtained via
the string theory / matrix model glueball potential $W_{\rm eff}(S_1, 
\dots S_K)$, upon integrating out the $S_i$, appears to always agree with
standard gauge theory results for the superpotential \refs{\ChoBI
\CsakiEU}, as seen in the examples of \CachazoKX\ and \ej.  This agreement
comes about via a remarkable interplay between the different terms $i$ in
\bk.  If we treated the scales $\Lambda _i$ of the $Sp(N_i)$ factors
as if they were initially independent, each term $\Nh_i \Pi (\ev{S_i})
-2\pi i \eta _i \ev{S_i}$ in \bk\ would be a complicated function of $\Lambda _i$,
which does not have a known, conventional, interpretation in terms of 
standard gauge theory for the low-energy $Sp(N_i)$ factor.  But upon adding
the different $i$ terms and using the matching relations relating $\Lambda _i$
to $\Lambda$, e.g. \laminsp, one nevertheless obtains the standard gauge
theory results, thanks to an intricate interplay between the different terms $i$.

By the map of \spasaux\ \CachazoKX, the agreement between string theory /
matrix models and standard gauge theory for $Sp(N)$ with antisymmetric can be phrased as such an
agreement for $U(N+K)$ with adjoint and breaking pattern
$U(N+K)\rightarrow \prod _{i=1}^KU(N_i+1)$.  

As another example, consider $U(N)$ with adjoint $\Phi$ and
superpotential having $K=N-1$, in the vacuum where
$U(N)\rightarrow U(2)\times U(1)^{N-2}$.  Factorizing the Seiberg-Witten
curve yields for the exact superpotential 
\ref\ElitzurGK{
S.~Elitzur, A.~Forge, A.~Giveon, K.~A.~Intriligator and E.~Rabinovici,
``Massless Monopoles Via Confining Phase Superpotentials,''
Phys.\ Lett.\ B {\bf 379}, 121 (1996)
[arXiv:hep-th/9603051].
} \eqn\wscph{W_{\rm exact}=W_{\rm cl}(g)\pm 2 g_{N}\Lambda ^N.}  The map
of \CachazoKX\ and sect.\ 3.3 relates this to $Sp(1)\rightarrow
Sp(1)\times Sp(0)^{N-2}$, where the exact gauge theory result agrees
with \wscph, up to the classical shift, upon using the relation \spausr.
A priori, one might expect the string theory / matrix model result to
disagree with \wscph, due to residual instanton contributions from the
$U(1)^{N-2}$ or the $Sp(0)^{N-2}$ in $U(N)\rightarrow U(2)\times
U(1)^{N-2}$ and $Sp(1)\rightarrow Sp(1)\times Sp(0)^{N-2}$ respectively.
But the string theory / matrix model result nevertheless agrees with
\wscph, thanks to the interplay between the different terms.  Consider,
in particular, the case $U(3)\rightarrow U(2)\times U(1)$.  The fact
that \wscph\ will only hold if remarkable cancellations occur upon
integrating out $S_1$ and $S_2$ from the non-trivial $W(S_1, S_2)$, was
discussed in \CachazoJY, where the cancellations were verified to indeed
occur, up to order $\alpha ^3$.  This is checked to one higher order in
\ej, since it is related to $Sp(1)\rightarrow Sp(1)\times Sp(0)$ by the
map of \CachazoKX\ and sect.\ 3.3.  The leading order cancellation, say
in terms of $U(3)\rightarrow U(2)\times U(1)$, is between $U(1)$ gaugino
condensation, $\Lambda _2^3$, and a higher order term coming from
integrating out $S_1$ from $W_{\rm pert}(S_i)$.

The residual instanton contributions associated with the UV completion \higgsbpsg, as
opposed to the standard gauge theory results for \higgsbp\ do not always cancel, however. 
 The cases where we find non-cancellations are when the
degree of the superpotential is sufficiently large, so that it contains terms which
are not independent moduli.  As an example, consider 
$U(1)$ with $W_{\rm tree}$ as in \ba\ having $K$ minima, breaking
$U(1)\rightarrow U(1)\times U(0)^{K-1}$.
The gaugino
condensation contribution to the superpotential, according to the string
theory (matrix model)
construction, is given by \wgca\  with $h_1=1$ and all other $h_i=0$ and
their $S_i$ set to zero.
Upon integrating out $S_1$, we thus obtain the superpotential
\eqn\wgczz{W_{\rm gc}=\Lambda _1^3=\Lambda ^2W''(a_1)=g_{K+1}\Lambda ^2\prod
_{j\neq 1}
(a_j-a_1),}
where we used the matching relation \lamin\ with $N_1=1$ and all other
$N_j=0$.

The full low-energy effective superpotential $W_{\rm low}(g_i, \Lambda)$ can
be regarded
as the generating function for the operator expectation values:
\eqn\wlowgf{\ev{u_j}={\partial W_{\rm low}(g_i, \Lambda)\over \partial g_j}
\qquad u_i\equiv {1\over j}
\Tr \Phi ^j.}  In the $U(1)$ theory, we have classical relations
$u_j={1\over j}u_1^j$.  But the
quantum contribution \wgczz\ (along with additional, higher order
contributions) imply
quantum deformation of these classical relations, due to the residual
instanton effects in the
$U(1+k|k)\rightarrow U(1+k_1|k_1)\times U(k_2|k_2)\dots U(k_K|k_K)$
F-completion.
For the simplest such example, consider $U(1)$ with $W_{\rm tree}=\half m
\Phi ^2 +\lambda
\Phi$.  The low-energy superpotential is
\eqn\wlowexq{W_{\rm low}=-{\lambda ^2\over 2m}+m\Lambda ^2,}
with the first term the classical contribution and the second the
residual instanton.
Using \wlowgf\ we then get
\eqn\evqis{\ev{u_1}=-{\lambda \over m}, \quad, \ev{u_2}={\lambda ^2\over
2m^2}+\Lambda ^2
\qquad\hbox{i.e.}\quad \ev{u_2}=\half \ev{u_1^2}+\Lambda ^2,}
which can be regarded as an instanton correction to the composite
operator $u_2$.

As another such example, consider $U(2)$ with an adjoint and $W_{\rm tree}$
having $K$
minima, in the vacuum where the gauge group is broken as $U(2)
\rightarrow U(1)\times U(1)\times U(0)^{K-2}$.  The gaugino condensation
contribution to $W_{\rm low}$ is
\eqn\wlowggf{W_{\rm gc}=\Lambda _1^3+\Lambda _2^3=g_{K+1}\Lambda
^4\left({\prod _{j=3}^K
(a_j-a_1)-\prod _{j=3}^K (a_j-a_2)\over a_2-a_1}\right).}
For example, for $U(2)$ with $W_{\rm tree}$ having $K=3$ critical points, we
break
$U(2)\rightarrow U(1)\times U(1)\times U(0)$ and \wlowggf\ leads to
\eqn\wlowcgexx{W_{\rm gc}=g_4\Lambda ^4.}
Computing expectation values as in \wlowgf\ this leads to
\eqn\uivc{\ev{u_4}=\ev{u_4}_{\rm cl}+\Lambda ^4,}
which can be interpreted as an instanton contribution to the composite
operator
$u_4=\Tr \Phi ^4$ in $U(2)$ gauge theory.  More generally, for $U(N)$
gauge theory,
the independent basis of operators $u_j={1\over j}\Tr \Phi ^j$ are only
those with $j\leq N$, those with $j>N$ can be expressed as products of
these basis operators via classical
relations.  But these relations can be affected by instantons.  In
particular, for $U(N)$
with an adjoint, the instanton factor is $\Lambda ^{2N}$, so operators
$u_j$ with $j\geq 2N$
can be affected.  The above residual instanton contributions of the
$U(N+k|k)$ UV completion can be interpreted as implying specific such
instanton corrections to the higher Casimirs
$u_j$.

A similar situation arises in the ${\cal N}=1^*$ $U(N)$ theory, where the
effective superpotential of the matrix model and conventional
gauge theory differ by a contribution $N^2m^3 E_2(N \tau )$ \DoreyTJ; this 
was interpreted in \DoreyTJ\ as differing operator definitions of $\Tr \Phi ^2$ between
gauge theory and the matrix model at the level of instantons.   Related issues for
multi-trace operators were seen in \BalaTM.

\newsec{Conclusions}

To compute the correct string theory / matrix model results, we should include or
not include the glueball fields $S_i$ according to the prescription of this paper.
Upon doing so, in all examples that we know of, the string theory / matrix model results 
agree with the results of standard gauge theory, at least in those cases where the
relevant gauge theory does not suffer from UV ambiguities.  In the case where such
ambiguities are present, for example in defining composite operators appearing in
$W_{\rm tree}$, the string theory / matrix model results correspond to a
particular UV definition of the theory.  The agreement with standard gauge theory
results is often due to a remarkable interplay between the different low-energy terms, 
found upon integrating out the glueball fields $S_i$, and connecting their scales 
$\Lambda _i$ via the appropriate
matching relation to the scale $\Lambda$ of the original theory.  In some cases, this
interplay leads to complete cancellations of the residual instanton contributions to $W_{\rm low}$ coming
{}from the $G(N+k|k)$ completion \refs{\DijkgraafXK, \AIVW}.  Perhaps there is 
some additional structure governing the glueball superpotentials, which would
make these remarkable  cancellations more manifest.

\bigskip\bigskip\bigskip
\leftline{\bf Acknowledgments}

\bigskip
We would like to thank M. Aganagic, F. Cachazo, S. Naculich and
H. Schnitzer for valuable discussions.

K.I. was supported in part by DOE-FG03-97ER40546 and KI would also like
to thank the Rutgers theory group for their support and hospitality
while he was a visitor, where part of this work was done.  P.K. was
supported in part by NSF grant PHY-0099570.  
The research of A.V.R was supported
by NSF grant 99-73935.  A.V.R. and C.V. 
thank the 2003 Simons Workshop in Mathematics and
Physics for a stimulating environment.
  C.V. would like to thank
the hospitality of theory group at Caltech, where he was a Distinguished
Gordon Moore Scholar.  The research of C.V. was supported in part by NSF
grants PHY-9802709 and DMS-0074329.

\appendix{A}{Matrix model calculation of superpotential}

In this appendix, after giving a proof for a general relation that
relates $S^2$ and $RP^2$ contributions to the $SO/Sp$ matrix
model free energy, we compute explicitly the free energy of the matrix
models associated with $SO/Sp$ gauge theory with adjoint
and $Sp$ gauge theory with antisymmetric tensor.  These matrix model
results are used in section 5 to evaluate the glueball
superpotential of the corresponding gauge theories.

\subsec{Proof for relation between $\CF_{S^2}$ and $\CF_{RP^2}$}

Here we prove a general relation between the $S^2$ and $RP^2$
contributions to the $SO(2N)/Sp(N)$ matrix model free energy:
\eqn\xaa{\eqalign{ \CF_{RP^2}= \cases{
 \displaystyle
 \mp{1\over2}{\partial \CF_{S^2}\over \partial S_0}
   & $SO(2N)/Sp(N)$ with adjoint,
 \cr\cr
 \displaystyle
 \mp{1\over2}\sum_{i=1}^K {\partial \CF_{S^2}\over \partial S_i}
   & $SO(2N)/Sp(N)$ with symmetric/antisymmetric tensor.
 }
}}
The first equation was conjectured in \refs{\ItaKX,\AshokBI} based on
explicit diagrammatic calculations, and proven in \JanikNZ\ for the case
of unbroken vacua.  Here we will give a general matrix model proof for
arbitrary breaking pattern.  These relations are equivalent to the maps
\soaaux, \spaaux, \spasaux, and \sosaux, which we obtained in sect.\ 3.3
immediately from the string theory geometric transition construction,
accounting for the orientifold contributions to the fluxes.

\bigskip
Consider $U(\mN)$ and $SO(2\mN)/Sp(\mN)$ matrix models which correspond
to $U(N)$ and $SO(2N)/Sp(N)$ gauge theories with a two-index tensor
matter field.  The partition function is
\eqn\xab{\eqalign{
 \mZ
 =
 e^{-\frac{1}{\mg^2}\mF(\mS_i)}
 =
 \int d\mPhi \,
   e^{-\frac{1}{\mg}W_{\rm tree}(\mPhi)}.
}}
We denote matrix model quantities by boldface letters, following the
notation of \KrausJV.  $\mPhi$ is an $\mN\times\mN$ (for $U(\mN)$
theory) or $2\mN\times 2\mN$ (for $SO(2\mN)/Sp(\mN)$ theory) matrix
corresponding to the $\Phi$ field in gauge theory, and the ``action''
$W_{\rm tree}$ is defined in \ba, \bsospx\ and \basp.
The matrix integral \xab\ is evaluated perturbatively around the general
broken vacua of \higgscases, with $N_i$ replaced by $\mN_i$.  We take
the double scaling limit $\mN_i\to\infty$, $\mg\to 0$ with $\mg
\mN_i\equiv\mS_i$ (for $U(\mN_i)$ factors) or $2\mg\mN_i\equiv \mS_i$
(for $SO(2\mN_i)/Sp(\mN_i)$ factors) kept finite.
The dependence of the free energy $\mF(\mS_i)$ on $\mN_i$ are eliminated
in favor of $\mS_i$, and $\mF(\mS_i)$ is expanded in the 't Hooft
expansion as
\eqn\xac{\eqalign{
 \mF(\mS_i)
 =\sum_{\CM} \mg^{2-\chi(\CM)} \CF_{\CM}(\mS_i)
 =\CF_{S^2}+\mg\CF_{RP^2}+\cdots
}}
where the sum is over all compact topologies $\CM$ of the matrix model
diagrams written in the 't Hooft double-line notation, and
$\chi(\CM)$ is the Euler number of $\CM$.

The matrix model resolvent is defined as follows:
\eqn\xad{\eqalign{
  \mR(z)
  \equiv
  \mg\Bracket{\Tr\left[\frac{1}{z-\mPhi}\right]}
  =
  \mR_{S^2}(z)+\mg\mR_{RP^2}(z)+\cdots.
}}
For $U(N)$ theory with adjoint, the expansion parameter is $\mg^2$
instead of $\mg$, and in particular, $\mR_{RP^2}(z)\equiv 0$.
The resolvent and the free energy are related as
\eqn\xadc{\eqalign{
 \mR_{\CM}(z)
 ={S\over z}\delta_{\chi(\CM),2}
 +{1\over z^2}{\partial\CF_{\CM}\over\partial g_1}
 +{2\over z^3}{\partial\CF_{\CM}\over\partial g_2}
 +{3\over z^4}{\partial\CF_{\CM}\over\partial g_3}
 +\cdots,
}}
where $S=\sum_{i=1}^K S_i$. The resolvents can be determined uniquely by
solving the matrix model loop equations (the loop
equations for the relevant matrix models are summarized in \KrausJV),
under the condition
\eqn\xada{\eqalign{ \oint_{A_i}{dz\over 2\pi i}\mR_{S^2}(z)=\mS_i,\qquad
\oint_{A_i}{dz\over 2\pi i}\mR_{RP^2}(z)=0. }}
$A_i$ is the contour around the $i$-th critical point of $W(z)$.  In
general, $\mR(z)$ develops a cut around each critical point
in the large $\mN_i$ limit, and $A_i$ is taken to encircle the $i$-th
cut. %
Note that the expression \xadc\ should be understood as a Laurent
expansion around $z=\infty$, and converges only if $|z|$ is
larger than $r$ such that all the singularities (cuts) of the resolvent
are inside the circle $C:~|z|=r$.

On the other hand, gauge theory resolvents $R(z)$, $T(z)$ (see e.g.\
\cdsw) are determined uniquely by solving the Konishi anomaly
equations (the Konishi anomaly equations for the relevant gauge theories
are summarized in \KrausJV), under the condition
\eqn\xadb{\eqalign{
  \oint_{A_i}{dz\over 2\pi i}R(z)=S_i,
  \qquad
  \oint_{A_i}{dz\over 2\pi i}T(z)=
    \cases{
      N_i & $U(N_i)$ \cr
      2N_i & $SO(2N_i)/Sp(N_i)$ \cr
    }.
}}

As was shown in \KrausJV, the matrix model resolvents
$\mR_{S^2}(z)$, $\mR_{RP^2}(z)$ are related to the gauge theory
resolvents $R(z)$, $T(z)$ as
\eqn\xae{\eqalign{
 R(z)=\mR_{S^2}(z),\qquad
 T(z)=
 \!\!\sum_{U(N_i)} \!\! 
   N_i{\partial\over\partial S_i}\mR_{S^2}(z)
 +\!\!\!\!\!\sum_{SO(2N_i)~\atop ~/Sp(N_i)} \!\!\!\!\!
   2N_i{\partial\over\partial S_i}\mR_{S^2}(z)
 +4\mR_{RP^2}(z)
}}
with $S_i$ and $\mS_i$ identified\foot{The relation \xae\ is an obvious
generalization of the formula in \KrausJV, which was for
unbroken vacua, to an arbitrary breaking pattern.  The gauge theory
resolvents $R(z)$, $T(z)$ given in \xae\ clearly satisfy the
condition \xadb\ provided that the matrix model resolvents
$\mR_{S^2}(z)$, $\mR_{RP^2}(z)$ satisfy the condition \xada.}.

\bigskip
First, consider $SO(2N)/Sp(N)$ theory with adjoint.  The general breaking
pattern is 
$SO(2N)\to SO(2N_0)\times U(N_1)\times \cdots \times U(N_K)$
or
$Sp(N)\to Sp(N_0)\times U(N_1)\times \cdots \times U(N_K)$
(Eq.\ \higgscases), where $N = N_0 + \sum_{i=1}^K
N_i$. Note that the eigenvalues are distributed in a
symmetric manner under $z\leftrightarrow -z$, and hence \xadb\ is
\eqn\xadea{\eqalign{ 
  \oint_{A_{0}}{dz\over 2\pi i}R(z)&=S_0,\qquad
  \oint_{A_{i}}{dz\over 2\pi i}R(z)
  =\oint_{A_{-i}}{dz\over 2\pi i}R(z)=S_i,
  \cr
  \oint_{A_{0}}{dz\over 2\pi i}T(z)&=2N_0,\qquad
  \oint_{A_{i}}{dz\over 2\pi i}T(z)
  =\oint_{A_{-i}}{dz\over 2\pi i}T(z)=N_i, 
}}
where $i=1,\dots,K$. The contours $A_i$ and $A_{-i}$ encircle
counterclockwise the cuts around $z=a_i$ and $z=-a_i$, respectively. The
relation \xae\ holds as it is, with the summation understood as
over $SO(2N_0)/Sp(N_0)$ and $U(N_i)$, $i=1,\dots,K$.

It was shown in \LandsteinerPH\ that the resolvents of this $SO(2N)/Sp(N)$
theory are related to the resolvents $\tilde R(z)$ and $\Tt(z)$
of $U(\Nt\!\equiv\!2N\mp 2)$ theory with adjoint as follows:
\eqn\xaf{\eqalign{
  R(z)=\tilde R(z),\qquad
  T(z)=\Tt(z)\pm{2\over z}.
}}
The tree level superpotential of the $U(\Nt)$ theory is related to the
one for the $SO(2N)/Sp(N)$ theory as $W^{U}(z)=W^{SO/Sp}(z)$ (see \ba\ and
\basp), and the breaking pattern is $U(\Nt)\to U(N_{-K})\times \cdots
\times U(N_{-1}) \times U(2N_0\mp 2) \times U(N_1)\times \cdots \times
U(N_K)$ with $N_{-i}=N_i$, $1\le i \le K$.  Note that since there is no
$z\leftrightarrow -z$ symmetry in the $U(\Nt)$ theory, the $U(N_{-i})$
factors that are ``images'' for $SO(2N)/Sp(N)$ are ``real'' for
$U(\Nt)$.  In addition, the glueball $\St$ of the $U(\Nt)$ theory is
related to the glueball $S$ of the $SO(2N)/Sp(N)$ theory as $\St_0=S_0$,
$\St_{i}=\St_{-i}=S_i$, $1\le i \le K$. Therefore, e.g.\ the first
equation in \xaf\ is more precisely
\eqn\xafa{\eqalign{
 R(z,S_j)
 =
 \Rt(z,\St_{j})\bigr|_{\St_0=S_0,\,\St_{i}=\St_{-i}=S_i}\,.
}}
Differentiating \xafa\ with respect to $S_j$, we obtain
\eqn\xafb{\eqalign{
  {\partial R\over\partial S_0} =
  {\partial \Rt\over \partial \St_0}
  \biggr|_{\St_0=S_0,\,~~~~\atop\St_i=\St_{-i}=S_i}
  ,\quad
  {\partial R\over\partial S_j} =
  \biggl(
  {\partial \Rt\over \partial \St_j}
  + {\partial \Rt\over \partial \St_{-j}}
  \biggr)
  \biggr|_{\St_0=S_0,\,~~~~\atop\St_i=\St_{-i}=S_i}
  ,
}}
where $1\le j\le K$.

Now, using \xae, let us translate the relation \xaf\ among gauge theory
resolvents into a relation among matrix model resolvents:
\eqn\xagg{\eqalign{
 \mR_{S^2}&=\tilde\mR_{S^2},\cr
 2N_0{\partial\mR_{S^2}\over\partial S_0}
 + \sum_{i=1}^K N_i {\partial\mR_{S^2}\over\partial S_i}
 + 4\mR_{RP^2}
 &=
 (2N_0\mp 2){\partial\tilde\mR_{S^2}\over\partial S_0}
 + \sum_{i=1}^K N_i
 \biggl(
  {\partial\tilde\mR_{S^2}\over \partial \St_i}
  + {\partial\tilde\mR_{S^2}\over \partial \St_{-i}}
 \biggr)
 \pm {2\over z}.
}}
Here $\tilde\mR_{S^2}$ is the matrix model resolvent associated with the
$U(\Nt)$ theory.  Using \xafb\ and the relations
$\mR_{S^2}=R$, $\tilde\mR_{S^2}=\Rt$ (Eq.\ \xae), we obtain
\eqn\xag{\eqalign{
 \mR_{RP^2}(z)
 =
 \mp {1\over 2} {\partial\over\partial S_0}\mR_{S^2}(z)
 \pm {1\over 2z}.
}}
By expanding the resolvents around $z=\infty$ using \xadc\ and comparing
the coefficients, we obtain a relation between matrix
model free energies:
\eqn\xaga{\eqalign{
 j{\partial \CF_{RP^2} \over \partial g_j}
 =
 \pm {1\over 2} {\partial\over \partial S_0}
 \left( S\delta_{j0}+  j{\partial \CF_{S^2}\over \partial g_j} \right)
\mp {1\over 2}\delta_{j0}.
}}
where $j=0,2,\cdots,2(K+1)$.  The $j=0$ case is trivially satisfied
since $S=S_0+2\sum_{i=1}^K S_i$ here, while the
$j=2,4,\dots,2(K+1)$ cases lead to the first equation of \xaa, which we
wanted to prove.

\bigskip
Next, consider $SO(2N)/Sp(N)$ theory with symmetric/antisymmetric
tensor.  The breaking pattern is $SO(2N)\to \prod_{i=1}^K SO(2N_i)$ or
$Sp(N)\to \prod_{i=1}^K Sp(N_i)$ (Eq.\ \higgscases), where
$N=\sum_{i=1}^K N_i$.  It was shown in \CachazoKX\ that the resolvents
of this $SO/Sp$ theory is related to the resolvents $\tilde R(z)$ and
$\Tt(z)$ of $U(\Nt\!\equiv\! 2N\mp 2K)$ theory with adjoint as follows:
\eqn\xai{\eqalign{
  R(z)=\tilde R(z),\qquad
  T(z)=\Tt(z)\pm{d\over dz}\ln[W'(z)^2+f_{K-1}(z)].
}}
The tree level superpotential of the $U(\Nt)$ theory is related to the
one for the $SO(2N)/Sp(N)$ theory as $W^{U}(z)=W^{SO/Sp}(z)$ (see \ba\
and \bsospx), and the breaking pattern is $U(\Nt)\to \prod_{i=1}^{K}
U(2N_i\mp 2)$.  The glueball $S_i$ of the $U(\Nt)$ theory is taken to be
the same as the glueball of the $SO(2N)/Sp(N)$ theory.  In \xai, 
$f_{K-1}(z)$ is a polynomial of degree $K-1$.  Using \xae, we can
translate the relation \xai\ among gauge theory resolvents into a
relation among matrix model resolvents:
\eqn\xaj{\eqalign{
 \mR_{RP^2}(z)
 =
 \mp {1\over 2} \sum_{i=1}^K {\partial\over\partial S_i}\mR_{S^2}(z) \pm
{1\over 4}{d\over dz}\ln[W'(z)^2+f_{K-1}(z)].
}}
In order to extract the relation between matrix model free energies, let
us multiply \xaj\ by $z^j$ ($0\le j\le K+1$) and
integrate over $z$ along the contour $C$, introduced under \xada, which
encloses all the cuts around the critical points of
$W(z)$.  Taking
\eqn\xak{\eqalign{
 W'(z)^2+f_{K-1}(z)
 =g_{K+1}^2\prod_{i=1}^K (z-a_i^+)(z-a_i^-),
}}
the branching points of the cuts are at $z=a_i^\pm$.  The second term on
the right hand side in \xaj\ does not contribute to the contour integral
unless $j=0$:
\eqn\xal{\eqalign{
 \mp &{1\over 4}\oint_C {dz\over 2\pi i}
 \sum_{i=1}^K z^j\left( {1\over z-a_i^+}+{1\over z-a_i^-}\right)\cr &=
 \pm {1\over 4}\oint_C {dw\over 2\pi i}
 \sum_{i=1}^K {1\over w^{j+1}}
 \left( {1\over 1-a_i^+w}+{1\over 1-a_i^-w}\right)
 =
 \mp {K\over 2}\delta_{j0},
}}
where $w=1/z$, because all the poles $w=1/a_i^{\pm}$ are outside of the
contour $C$ (on the $w$-plane).  On the contour $C$, we can use the
Laurent expansion \xadc\ to evaluate the contribution from the other
terms, and the final result is
\eqn\xam{\eqalign{
 j{\partial \CF_{RP^2} \over \partial g_j}
 =
 \pm {1\over 2}\sum_{i=1}^K {\partial\over \partial S_i}
 \left( S\delta_{j0}+  j{\partial \CF_{S^2}\over \partial g_j} \right)
\mp {K\over 2}\delta_{j0}.
}}
The $j=0$ case is trivially satisfied, while the $1\le j\le K+1$ cases
lead to the second equation of \xaa, which we wanted to
prove.

\subsec{Computation of matrix model free energy: $SO(2N)/Sp(N)$ theory
with adjoint}

Let us consider $SO(2\mN)$ matrix model which corresponds to $SO(2N)$
gauge theory with adjoint.  The tree level superpotential is
taken to be quartic \da.  The matrix variable $\mPhi$ in \xab\ is a real
antisymmetric matrix and can be skew-diagonalized as
\eqn\xb{\eqalign{
  \mPhi \cong
  {\rm diag}[\lambda_1,\cdots,\lambda_{\mN}]\otimes i\sigma^2.
}}
By changing the integration variables from $\mPhi$ to $\lambda_i$, we
obtain %
\eqn\xc{\eqalign{
  \mZ\sim\int
  \prod_{i=1}^{\mN}d\lambda_i
  \prod_{i<j}^{\mN}(\lambda_i^2-\lambda_j^2)^2\,\,
  e^{-{1\over \mg}\sum_{i=1}^{\mN}
  \left(-{m\over 2}\lambda_i^2+{g\over 4}\lambda_i^4\right)},
}}
where $\prod_{i<j}^{\mN}(\lambda_i^2-\lambda_j^2)^2$ is the Jacobian
for this change of variables \refs{\Mehta,\AshokBI}.  The
polynomial $-{m\over 2}\lambda^2+{g\over 4}\lambda^4$ has critical
points at $\lambda=0,\pm\sqrt{m/g}$, around which we would like
to do perturbative expansion.  For this purpose, we separate $\lambda$'s
into two groups as
\eqn\xf{\eqalign{
  \lambda_i =
  \cases{
    \lambda_{i_0}^{(0)}            & $i_0=1,\dots,\mN_0$,\cr
    \sqrt{m/g}+\lambda_{i_1}^{(1)} & $i_1=1,\dots,\mN_1$,
  }
}}
with $\mN_0+\mN_1=\mN$, corresponding to the classical
supersymmetric vacuum with breaking pattern $SO(2\mN ) \to SO(2\mN _0)\times
U(\mN _1)$.
%
We would like to evaluate the matrix integral \xc\ perturbatively around
$\lambda^{(0,1)}=0$.
If we expand the matrix model free energy in the coupling constant $g$
as %
\eqn\xi{\eqalign{ \mF =g f_1(\mN_0,\mN_1)+ g^2 f_2(\mN_0,\mN_1)+\cdots,
}} %
the loop expansion tells us that $f_n(\mN_0,\mN_1)$ is a polynomial of
degree $n+2$.  Therefore, by performing the matrix integral by
computer for small values of $\mN_0$ and $\mN_1$, one can determine the
polynomial $f_n$.  If we rewrite $\mN_{0,1}$ in favor of
$S_{0}=2\mg \mN_{0}$ and $S_{1}=\mg \mN_{1}$,
the expansion \xi\ arranges itself into the 't Hooft expansion \xac,
from which one can read off $\CF_{S^2}$, $\CF_{RP^2}$, etc.

Following the procedure sketched above, we computed the matrix model
free energy as
\eqn\xk{\eqalign{
  \CF_{S^2}=&
  \left( {1\over 4}S_0^3-2 S_0^2 S_1+ S_0 S_1^2 \right)\alpha
  +
  \left(
    -{9\over 16}S_0^4 + 7S_0^3 S_1 - {9}S_0^2 S_1^2+ {2}S_0 S_1^3
  \right)\alpha^2
  \cr
  &
  +
  \left(
    {9\over 4} S_0^5
    -{233\over 6} S_0^4 S_1
    +{262\over 3} S_0^3 S_1^2
    -{152\over 3} S_0^2 S_1^3
    +{20\over 3} S_0 S_1^4
  \right)\alpha^3
  +\CO(\alpha^4),\cr
}}
where we defined $\alpha\equiv{g/m^2}$.  We also checked explicitly
that the relation \xaa\ holds.  Substituting \xk\ into the DV
relation \dc, we obtain the superpotential \dd.

The $Sp(N)$ result is obtained similarly, with the result as in
\xaa.

\subsec{Computation of matrix model free energy: $Sp(N)$ theory with
antisymmetric tensor}

Consider the $Sp(\mN)$ matrix model which corresponds to $Sp(N)$ gauge
theory with an antisymmetric tensor.  The superpotential is
taken to be quartic \ea.  The matrix variable $\mPhi$ satisfies
$\mPhi=\mA J$, $\mA^T=-\mA$.
The ``action'' $W_{\rm tree}$ is given in \ea.
By a complexified $Sp(\mN)$ gauge rotation, the matrix $\mPhi$ can be
brought to the form \ChoBI
\eqn\yb{\eqalign{
  \Phi \cong {\rm diag}[\lambda_1,\dots, \lambda_{\mN}]
  \otimes{\bf 1}_2, \qquad
  \lambda_i\in {\bf C}.
}}
By changing the integration variables from $\Phi$ to $\lambda_i$, we
obtain %
\eqn\yc{\eqalign{
  Z\sim
  \int
    \prod_{i=1}^{\mN}d\lambda_i
    \prod_{i<j}^{\mN}(\lambda_i-\lambda_j)^4\,\,
    e^{
        -{1\over \mg}\sum_{i=1}^{\mN}
        \left( {m\over 2}\lambda_i^2+{g\over 3}\lambda_i^3 \right)
      }.
}}
where $\prod_{i<j}^{\mN}(\lambda_i-\lambda_j)^4$ comes from the Jacobian
for this change of variables.  The polynomial ${m\over
2}\lambda^2+{g\over 3}\lambda^3$ has two critical points $z=0,-{m\over
g}$, around which we would like to do perturbative
expansion.  For this purpose, we separate $\lambda$'s into two groups as
\eqn\yd{\eqalign{
  \lambda_i =
  \cases{
    \lambda_{i_0}^{(1)}      & $i_1=1,\dots,\mN_1$,\cr
    -m/g+\lambda_{i_1}^{(2)} & $i_2=1,\dots,\mN_2$,
  }
}}
with $\mN_1+\mN_2=\mN$.  This corresponds to the classical
supersymmetric vacuum with breaking pattern $Sp(\mN ) \to Sp(\mN_1)\times
Sp(\mN_2)$.

The matrix integral can be performed just the same way as for the
$SO/Sp(N)$ theory with adjoint, as described in the last
subsection. After substitution $S_{1,2}=2\mg \mN_{1,2}$, we obtain
\eqn\yg{\eqalign{ \CF_{S^2}=&
 \left(
  -{S_1^3\over 3}+{S_2^3\over 3}
  +{5\over 2}S_1^2 S_2 -{5\over 2}S_1 S_2^2
 \right)\alpha
\cr &+
 \left(
  -{4\over 3}S_1^4 + {91\over 6}S_1^3 S_2 - {59\over 2}S_1^2 S_2^2 +
{91\over 6}S_1 S_2^3-{4\over 3}S_2^4
 \right)\alpha^2
\cr &+
 \left(
  -{28\over 3}S_1^5
  + {871\over 6}S_1^4 S_2
  - {1318\over 3}S_1^3 S_2^2
  + {1318\over 3}S_1^2 S_2^3
  - {871\over 6}S_1 S_2^4
  +{28\over 3}S_2^5
 \right)\alpha^3
+\CO(\alpha^4),\cr
}}
where $\alpha\equiv g^2/m^3$.  We also checked explicitly that the
relation \xaa\ holds. Substituting \yg\ into \ed, we obtain
the glueball superpotential \eg.

\appendix{B}{Gauge theory calculation of superpotential}

In this appendix, we compute the exact superpotential of the $\CN=1$
$SO(2N)/Sp(N)$ theory with adjoint in various vacua by considering 
factorization of the $\CN=2$ curve.
This factorization method was developed in \CachazoJY\ for $U(N)$, and
generalized in \CachazoZK\ to the case with unoccupied critical points
(in other words, the $n<K$ case below). Inclusion of fundamentals was
considered in \BalasubramanianTV.  The generalization to $SO/Sp$ gauge group,
which discuss below, was given in
\refs{\EdelsteinMW,\FujiVV,\FengGB,\AhnCQ,\AhnVH,\AhnUI}.

First consider $\CN=2$ $SO(2N)$ theory broken to $\CN=1$ by the following
polynomial tree level superpotential for the adjoint
chiral superfield $\Phi$:
\eqn\za{\eqalign{ 
  W_{\rm tree}&=\half\Tr[W(\Phi)], \cr 
  W(x) &= \sum_{j=1}^{K+1}{g_{2j}\over 2j}x^{2j},\qquad
  W'(x)=g_{2K+2}\,x\prod_{i=1}^{K}(x^2-a_i^2). 
}}
The classical supersymmetric vacua are obtained by putting $2N_0$
eigenvalues of $\Phi$ at $x=0$ and $N_i$ pairs of eigenvalues at
$x=\pm a_i$, where $i=1,\cdots,K$ and $N_0+\sum_{i=1}^K N_i=N$.  In
this vacuum the gauge group breaks as $SO(2N)\to SO(2N_0)\times
\prod_{i=1}^K U(N_i)$.  We allow some of $N_i$ to vanish, i.e.\ we allow
``unoccupied'' critical points.  Let the number of
nonzero $N_{i\ge 1}$ be $n$.  Then, the $\CN=2$ curve governing this
$SO(2N)$ theory factorizes as
\refs{\EdelsteinMW,\FujiVV,\FengGB}:
\eqn\zb{\eqalign{ 
  y^2=P_{2N}^2(x)-4x^4\Lambda^{4N-4}
  = \cases{
      [xH_{2N-2n-2}(x)]^2 F_{2(2n+1)}(x) & $N_0> 0$,\cr 
      H_{2N-2n}(x)^2 F_{4n}(x) & $N_0=0$. 
    } 
}}
Here $P$, $H$ and $F$ are polynomials in $x$ of the subscripted degree, 
which are invariant under $x\to -x$, i.e.\ they are
actually polynomials in $x^2$.  This factorization is required to have the 
appropriate number of independent, massless, monopoles and dyons, which
must condense to eliminate some of the low-energy photons.  The polynomial $F$ 
is related to the
tree level superpotential as
\eqn\zc{\eqalign{\cases{ F_{2(2n+1)}(x)={1\over
g_{2K+2}^2}W'_{2K+1}(x)^2+f_{2K}(x)  &   $n=K$, \cr
F_{2(2n+1)}(x)Q_{2K-2n}(x)^2={1\over g_{2K+2}^2}W'_{2K+1}(x)^2+f_{2K}(x)
& $n<K$ } }}
with some polynomial $Q_{2K-2n}(x)$, $f_{2K}(x)$ of the subscripted
degrees, and $W'_{2K+1}(x)$ is as in \Wprimeisp.
Equation \zc\ is for $N_0>0$, and for $N_0=0$ one must use the second
equation with $Q_{2K-2n}$ replaced by $Q_{2K-2n+2}$.

For $N_0>0$ we can write the solution of \zb\ in terms of that of the corresponding
$U(2N-2)$ breaking pattern, via:
\eqn\psopu{P^{SO(2N)}_{2N}(x)=x^2P_{2N-2}^{U(2N-2)}(x).}

The low energy superpotential is given by 
\eqn\ze{\eqalign{ 
  W_{\rm low} = \half\sum_{j=1}^{K+1}g_{2j}\ev{u_{2j}}, 
}}
where the $\ev{u_{2j}}$ are constrained to satisfy \zb.  Implementing this 
leads to the result that  \AhnCQ:
\eqn\zd{\eqalign{ 
  \left\langle\Tr{1\over x-\Phi}\right\rangle 
  = {d\over dx}\ln 
  \left[P_{2N}(x)+\sqrt{P_{2N}(x)^2-4x^4\Lambda^{4N-4}}\right].
}}
%
Plugging back into \ze\ gives $W_{\rm low}$.  
Note that the superpotential takes this simple form \ze\ only after one
integrates out the monopoles and dyons, whose equation of
motion led to the factorization constraint \zb\ \refs{\SeibergRS
,\SeibergAJ ,\JanikNZ}.

The $Sp(N)$ theory can be solved similarly.
The $\CN=2$ curve factorizes in the vacuum with breaking
pattern $Sp(N)\to Sp(N_0)\times \prod_{i=1}^K U(N_i)$ as
\refs{\EdelsteinMW,\FujiVV,\FengGB}
\eqn\zf{\eqalign{ 
  y^2&=B_{2N+2}(x)^2-4\Lambda^{4N+4}= x^2H_{2N-2n}(x)^2  F_{2(2n+1)}(x), \cr 
  B_{2N+2}(x)&\equiv x^2P_{2N}(x)+2\Lambda^{2N+2}. 
}}
The polynomial $F_{2(2n+1)}(x)$ is related to $W'(x)$ by \zc.
The mapping of the $Sp(N)$ theory to a $U(2N+2)$ theory, as in \spaaux,
can be written as a solution of \zf\ in terms of solutions of the corresponding
$U(2N+2)$ factorization problem:
\eqn\psppu{B_{2n+2}(x)\equiv x^2P_{2N}^{Sp(N)}(x)+2\Lambda ^{2N+2}=P_{2N+2}^{U(2N+2)}(x).}  
The $\Lambda ^{2N+2}$ shift in \psppu\ is an $Sp(N)$ residual instanton effect,
associated with the index of the embedding of the $U(N_i)$ factors
in $Sp(N)$ \refs{\IntriligatorID, \CsakiVV}

Again, the superpotential is given as in \ze, subject to the constraint that 
$\ev{u_j}$ satisfy \zf.  Implementing this, the $\ev{u_j}$ can be obtained
{}from $B_{2N+2}(x)$ by \AhnCQ
\eqn\zg{\eqalign{
  \left\langle\Tr{1\over x-\Phi}\right\rangle 
  = {d\over dx}\ln  \left[B_{2N+2}(x)+\sqrt{B_{2N+2}(x)^2-4\Lambda^{4N+4}}\right]. 
}}

We now consider the exact $W_{\rm eff}$ for a few $SO/Sp(N)$ 
cases, to illustrate and clarify the general
features\foot{More $SO/Sp$ examples can be found in \AhnCQ.}.  We
take quartic tree level superpotential
\eqn\zh{\eqalign{W_{\rm tree}=\half \Tr W(\Phi), \qquad  W(x)={m\over 2}x^2+{g\over 4}x^4, }}
which corresponds to $K=1$.

\subsec{$SO(2N)$ unbroken}

By the map \soaaux, this maps to $U(2N-2)$ unbroken, for which $P^{U(2N-2)}(x)=
2\Lambda ^{2N-2}T_{2N-2}(x/2\Lambda)$, with $T_N(x=\half(t+t^{-1}))=\half (t^N+t^{-N})$ a Chebyshev polynomial \DouglasNW.  Thus, using \psopu, $P^{SO(2N)}(x)=2\Lambda ^{2N-2}x^2T_{2N-2}(x/2\Lambda)$, as found in \JanikNZ.   This then leads to \JanikNZ
\eqn\sounbrv{\ev{u_{2p}}\equiv {1\over 2p} \ev{\Tr \Phi ^{2p}}={2N-2\over 2p}{\pmatrix{2p\cr p}}
\Lambda ^{2p}.}
In particular, 
\eqn\zk{\eqalign{\ev{u_2}=(2N-2)\Lambda^2,\quad \ev{u_4}=3(N-1)\Lambda^4,}}
and the low-energy superpotential is $W_{\rm low}=\half (m\ev{u_2}+g\ev{u_4})$:
\eqn\zl{\eqalign{ W_{\rm low}=(N-1)\left( m\Lambda^2+{3\over 2}g\Lambda^4\right). }}

\subsec{$SO(2N)\to SO(2)\times U(N-1)$}

By the map \soaaux, this maps to $U(2N-2)\rightarrow U(0)\times U(N-1)\times U(N-1)$.
Using \soaaux, the multiplication map of \CachazoJY\ for the $U(2N-2)$ theory leads to
a similar multiplication map 
for the $SO(2N)$ theory, which was discussed in \AhnCQ.  Using this,
we can construct the solution to the factorization problem for general $N$ in terms of that
of say $N=2$, i.e.\ $SO(4)\rightarrow SO(2)\times U(1)$.  In this case, 
equation \zb\ is
\eqn\zm{\eqalign{ y^2=P_4^2-4x^4 \Lambda^{4} = x^2 F_6. }}
The solution to this factorization problem is
\eqn\zn{\eqalign{ P_4=x^2(x^2-a^2),\quad
F_6=x^2[(x^2-a^2)^2-4\Lambda^2], }} %
from which we can see the breaking pattern $SO(4)\to SO(2)\times U(1)$.  Using
\zd\ gives
\eqn\zo{\eqalign{ u_2=a^2,\quad u_4={a^4\over 2}+\Lambda^4. }}
Further, the condition \zc\
\eqn\zp{\eqalign{ F_6={1\over g^2}W_3'{}^2+f_2 }}
leads to
\eqn\zq{\eqalign{ a^2=-{m\over g},\quad f_2=-4\Lambda^2 x^2. }}
The solution for general $SO(2N)\rightarrow SO(2)\times U(N-1)$, the multiplication map gives the solution to the factorization problem as $P_{2N}(x)=2x^2 \Lambda ^{2N-2}T_{N-1}((x^2-a^2)/2\Lambda ^2)$, with $T_{N-1}$ the Chebyshev polynomial defined above.  The effect is to 
rescale $u_2$, $u_4$, and hence $W_{\rm low}$ by an overall factor of $N-1$:
\eqn\zr{\eqalign{ W_{\rm low}=(N-1)\left(-{m^2\over 4g}+{1\over 2}g\Lambda^4\right). }}
This agrees with the result \di.  
\subsec{$SO(4)\to U(2)$}

More generally, we could consider the breaking pattern $SO(2N)\rightarrow U(N)$.  The
map of \soaaux\ is less useful here, when $N_0=0$, since it suggests mapping to
$U(2N-2)\rightarrow U(-2)\times U(N)\times U(N)$ and the $U(-2)$ needs to be interpreted.
In general, this breaking pattern leads to a complicated $W_{\rm low}(\Lambda)$.  We will here 
illustrate the case $SO(4)\to U(2)$, corresponding to 
$N=2,~N_0=0,~n=1,~K=1$. Equation \zb\ is
\eqn\zs{\eqalign{ y^2=P_4^2-4x^4 \Lambda^{4} = H_2^2 F_4. }}
The solution to this factorization problem is
\eqn\zt{\eqalign{ P_4=(x^2-a^2)^2+2\Lambda^2x^2,\quad H_2=x^2-a^2,\quad
F_4=(x^2-a^2)^2+4\Lambda^2x^2. }}
In the classical $\Lambda\to 0$ limit, this shows $P_4\to (x^2-a^2)^2$,
implying the breaking pattern $SO(4)\to U(2)$.  \zd\ gives
\eqn\zu{\eqalign{ u_2=2(a^2-\Lambda^2),\quad
u_4=(a^4-2\Lambda^2)^2-\Lambda^4. }}
Further, the condition \zc\
\eqn\zv{\eqalign{ F_4x^2={1\over g^2}W_3'{}^2+f_2 }}
leads to
\eqn\zw{\eqalign{ 
a^2=-{m\over g}+2\Lambda^2,\quad
f_2=4\Lambda^2x^2\Bigl(-{m\over g}+\Lambda^2\Bigr). 
}}
Therefore the exact superpotential is
\eqn\zx{\eqalign{ W_{\rm low} = -{m^2\over 2g}+m\Lambda^2+{1\over 2}g\Lambda^4. }}

\subsec{$Sp(2)\to U(2),\, Sp(1)\times U(1)$} 

This corresponds to
$N=2,~n=1,~K=1$. Equations \zf\ and \zc\ are
\eqn\zy{\eqalign{ y^2=B_6^2-4\Lambda^{12} = x^2H_2^2 F_6,\qquad
F_6={1\over g^2}W_3'{}^2+f_2. }}
This factorization problem is solved by \AhnCQ:
\eqn\zz{\eqalign{ P_4=(x^2-a^2)^2+{4\Lambda^6\over a^4}(x^2-2a^2),
\qquad {m\over g}=-a^2+{4\Lambda^6\over a^4} }}
From \zg, we obtain
\eqn\zaa{\eqalign{ u_2=2a^2-{4\Lambda^6\over a^4},\qquad
u_4=a^4+{8\Lambda^{12}\over a^8}. }}
This solution continuously connects two classically different vacua with
breaking pattern $Sp(2)\to U(2)$ and $Sp(2)\to
Sp(1)\times U(1)$. Correspondingly there are two ways to take the
classical limit: i) $\Lambda\to 0$ with $a$ fixed, and ii)
$\Lambda,a\to 0$ with $w=2\Lambda^3/a^2$ fixed.  In these limits,
$P_4(x)$ goes to i) $(x^2-a^2)^2$ and ii) $x^2(x^2+w^2)$,
showing the aforementioned breaking pattern.

In the $Sp(2)\to U(2)$ case, we solve the second equation of \zz\ with
the condition $a^2\to -m/g$ as $\Lambda\to 0$.  The
solution  is
\eqn\zab{\eqalign{ 
  a^2=-{m \over g}+{4g^2\Lambda^6\over
m^2}+{32g^5\Lambda^{12}\over m^5}+ {448g^8\Lambda^{18}\over m^8}+\cdots.
}}
From \zaa\ and \ze, one obtains the exact superpotential:
\eqn\zac{\eqalign{ 
  W_{\rm low}
  =
  -{m^2\over 2g}
  -{2g^2\Lambda^6\over m}
  -{4g^5\Lambda^{12}\over m^4} 
  -{32g^8\Lambda^{18}\over m^7}
  +\cdots. 
}}

In the $Sp(2)\to Sp(1)\times U(1)$ case, we solve the second equation of
\zz\ with the condition $w^2\to m/g$ as $\Lambda\to 0$.
It is
\eqn\zad{\eqalign{ w={m^{1/2}\over g^{1/2}}+{g\Lambda^3\over m}
-{3g^{5/2}\Lambda^6\over 2m^{5/2}}+{4g^4\Lambda^9\over
m^4}+\cdots. }}
From \zaa\ and \ze, one obtains the exact superpotential:
\eqn\zae{\eqalign{ 
  W_{\rm low}
  =
  -{m^2\over 4g}
  +2m^{1/2}g^{1/2}\Lambda^3
  +{g^2\Lambda^6\over m} 
  -{g^{7/2}\Lambda^9\over m^{5/2}}
  +\cdots. 
}}

\listrefs
\end